\documentclass[a4paper,11pt]{article}
\usepackage{jcappub} % for details on the use of the package, please
\usepackage[T1]{fontenc} % if needed
\usepackage{natbib}
\usepackage{amssymb}
\usepackage{amsmath}
\usepackage{amsfonts}
\usepackage{graphicx}
\usepackage[hang,tight,raggedright]{subfigure}
\usepackage{multirow}
\usepackage{subfigure}
\usepackage{float}
\usepackage{graphics}
\usepackage{slashed}
\usepackage{color}
\usepackage{bm}
\usepackage{braket}
\usepackage{psfrag}
\usepackage{booktabs}
\usepackage{latexsym}
\usepackage{pythonhighlight}
\newcommand{\be}{\begin{equation}}
\newcommand{\ee}{\end{equation}}
\newcommand{\besp}{\begin{equation}\begin{split}}
\newcommand{\eesp}{\end{split}\end{equation}}

\title{Model-dependent analysis method for energy budget of the cosmological first-order phase transition}

\author[a]{Xiao Wang,}
\author[b,1]{Chi Tian\note{Corresponding author.},}
\author[a]{and Fa Peng Huang}

\affiliation[a]{MOE Key Laboratory of TianQin Mission, TianQin Research Center for Gravitational Physics $\&$ School of Physics and Astronomy, Frontiers Science Center for TianQin, CNSA Research Center for Gravitational Waves, Sun Yat-sen University (Zhuhai Campus), Zhuhai 519082, China}
\affiliation[b]{School of Physics and Optoelectronics Engineering, Anhui University, 111 Jiulong Road, Hefei, Anhui 230601, China}

\emailAdd{wangxiao7@mail.sysu.edu.cn}
\emailAdd{ctian@ahu.edu.cn}
\emailAdd{huangfp8@mail.sysu.edu.cn}

\abstract{The kinetic energy of the fluid shell in the cosmological first-order phase transition is crucial for predicting the gravitational wave signals generated by the sound wave mechanism.
We propose a model-dependent method to calculate the kinetic energy fraction by dividing the bubble-fluid system into three distinct regions: the symmetric phase, the broken phase, and the bubble wall. 
By solving the local equation of motion of the scalar field with a phenomenological friction term, the bubble wall velocity and the boundary conditions of the fluid equations of both phases can be derived simultaneously. 
Then, for a given particle physics model, the fluid profiles of different hydrodynamical modes and the corresponding kinetic energy fraction can be obtained. 
Our method can also capture the temperature dependency of the sound speed of the plasma. 
Compared with the conventional model-independent method, our approach is based on an accurate equation of state derived directly from the effective potential and takes into account the contribution of the bubble wall to the energy-momentum tensor.
Therefore, our method in-principle provides a more consistent and accurate result, which is crucial for  high-precision calculations of the gravitational waves induced by the first-order phase transition.
}

\begin{document}
	\maketitle
	\flushbottom
%	\newpage

\section{Introduction}

The cosmological first-order phase transition (FOPT) and its associated phase transition gravitational waves (GWs) open a new window to explore many fundamental problems in particle cosmology, such as electroweak baryogenesis~\cite{Trodden:1998ym,Morrissey:2012db} and dark matter~\cite{Baker:2019ndr,Chway:2019kft,Huang:2017kzu,Huang:2017rzf,Elor:2021swj}, etc.
During an FOPT, GW signals can be generated by bubble collisions~\cite{Kosowsky:1991ua,Kosowsky:1992vn,Huber:2008hg}, sound waves~\cite{Hindmarsh:2013xza,Hindmarsh:2015qta,Hindmarsh:2017gnf}, and turbulence~\cite{Kosowsky:2001xp,Caprini:2009yp,RoperPol:2019wvy}.
Future GW experiments, such as TianQin~\cite{TianQin:2015yph,TianQin:2020hid}, LISA~\cite{LISA:2017pwj}, Taiji~\cite{Hu:2017mde}, and Big Bang Observatory (BBO)~\cite{Corbin:2005ny}, among others, may be able to detect phase transition GW signals and reveal various puzzles about our Universe.
Recent studies~\cite{Hindmarsh:2013xza,Hindmarsh:2015qta,Hindmarsh:2017gnf} have shown that the sound wave is the dominant source of phase transition GWs in a thermal FOPT.
To finally pin down the underlying physics of an FOPT, we need precise quantification of the GW spectra and their signal-to-noise ratio in future GW experiments.
To that end, obtaining accurate estimations of phase transition parameters that determine the GW spectra become critical.
The two most crucial variables among these parameters are the bubble wall velocity and the energy budget, where the latter is also known as the kinetic energy fraction, which characterizes the fraction of energy released during an FOPT that is transferred into the kinetic energy of the ambient fluid.

Previous studies~\cite{Espinosa:2010hh,Leitao:2010yw,Leitao:2014pda,Giese:2020rtr,Giese:2020znk,Wang:2020nzm,Wang:2022lyd} on the energy budget of an FOPT are based on the conventional model-independent method. This method matches a well-motivated particle physics model to a specific model of equation of state (EoS) according to some phase transition parameters, such as the strength parameter and the bubble wall velocity. 
As a result, the boundary conditions of the fluid equations can be determined, and then the corresponding kinetic energy fraction can be derived. 
However, the specific model of EoS, such as the bag model, cannot fully capture the features of the symmetric and broken phases since some important factors are ignored, such as the variation of the sound speed of the plasma.
To address these limitations, some other models of EoS are constructed. For example, the $\mu\nu$ model~\cite{Leitao:2014pda,Giese:2020rtr,Giese:2020znk,Wang:2020nzm} can take the variation of the sound speed into account by considering it as an extra parameter. Additionally, to incorporate the temperature-dependent sound speed, a more complicated model of EoS is built in ref.~\cite{Wang:2022lyd}, which includes more extra parameters. 
Although the model-independent method is convenient for studying various particle physics models, the accuracy of this method depends on the constructed EoS model. 
To capture all relevant properties, a complicated EoS model should be considered, and sophisticated calculations could be involved. Furthermore, the calculation of the bubble wall velocity is decoupled from the kinetic energy fraction in the conventional model-independent method.%, which might not always be the case.

%Recent studies~\cite{Balaji:2020yrx,Ai:2021kak,Laurent:2022jrs,Wang:2022txy} have shown that the fluid dynamics can have a significant impact on the bubble wall velocity, which is strongly correlated with the the energy budget.
The bubble wall velocity is strongly correlated with the energy budget, and
recent studies~\cite{Balaji:2020yrx,Ai:2021kak,Laurent:2022jrs,Wang:2022txy} show that the fluid dynamics can have a significant impact on it.
To this end, in this work, we unify the calculation of the energy budget and the bubble wall velocity into a single framework.
Given that the phase transition system is a coupled fluid-scalar system, we divide it into three regions, which are the symmetric phase, the broken phase, and the bubble wall. 
The plasma in the symmetric and broken phases is treated as a perfect fluid, and the corresponding EoS of both phases are directly obtained from the effective potential of the particle physics model, which is actually the free energy of the phase transition system. 
In contrast to the conventional model-independent method, our approach does not require the construction of a model of EoS  but derives it directly from the effective potential.
To obtain the kinetic energy fraction, the fluid equations in both phases with proper boundary conditions need to be solved. In our framework, we derive the corresponding boundary conditions by solving the local equations of motion (EoM) of the scalar field and the fluid across the bubble wall. 
To simplify the calculation, a phenomenological friction term characterizing the out-of-equilibrium phenomena is introduced. 
With the planar approximation, we solve the scalar field and fluid profiles, thereby obtaining the bubble wall velocity and the boundary conditions for the fluid equations of both phases. This allows us to finally solve the fluid profiles in both phases and derive the energy budget.
Since the contribution of the scalar field to the energy-momentum tensor is taken into account, and the EoS is directly obtained from the effective potential,
our model-dependent method can provide more consistent and accurate results in principle.

This work is organized as follow.
In section~\ref{sec:hydro}, we introduce the general hydrodynamical treatment of the FOPT system. 
Then we propose our model-dependent method for the calculation of energy budget in section~\ref{sec:Kdefdet}, where the definition of the kinetic energy fraction and a review of the conventional model-independent method to quantify it are also given. 
In section~\ref{sec:model}, we use a representative model to demonstrate our model-dependent analysis method. 
Discussion and conclusion are given in section~\ref{sec:dic} and section~\ref{sec:con}, respectively.

\section{Hydrodynamics}
\label{sec:hydro}
The thermal cosmological FOPT system consists of scalar fields, which act as the order parameter of the phase transition, and the thermal plasma.
The energy-momentum tensor of $N$ scalar fields is
\begin{equation}
T_{\phi}^{\mu\nu} = \partial^\mu\phi_i\partial^\nu\phi_i - g^{\mu\nu}\left[\frac{1}{2}\partial_{\alpha}\phi_i\partial^{\alpha}\phi_i - V_0(\Phi)\right], \label{eq:EMphi}
\end{equation}
where $V_0(\Phi)$ is the one-loop zero temperature potential,  
the repeated indices are summed over and $\Phi\equiv\{\phi_1, \phi_2, ...,\phi_N \}$.
The energy-momentum tensor of the plasma can be derived by
\begin{equation}
T_{\rm pl}^{\mu\nu} = \sum_i\int\frac{d^3k}{(2\pi)^3E_i}k^\mu k^\nu f_i(k, x),
\end{equation}
where $f_i(k,x)$ is the distribution function for different particle species and the sum is over all particle species in the plasma.
If we assume the plasma is perfect fluid, the corresponding energy-momentum tensor can be given by
\begin{equation}
	T_{\rm pl}^{\mu\nu} = (e + p)u^\mu u^\nu - g^{\mu\nu}p,\label{eq:EMpl}
\end{equation}
where $e$ and $p$ are the energy and pressure of the plasma, respectively,
and the four-velocity $u^\mu$ is 
\begin{equation}
u^{\mu} = \frac{1}{\sqrt{1 - \mathbf{v}^2}}(1,\mathbf{v}) = (\gamma, \gamma\mathbf{v}),
\end{equation}
where $\mathbf{v}$ is the three-velocity and $\gamma$ is the Lorentz factor.
Note that the pressure $p$ also contains the contribution of the scalar fields.
And the relations of these thermodynamical quantities are
\begin{equation}
w \equiv T\frac{\partial p}{\partial T}, \quad e\equiv T\frac{\partial p}{\partial T} -p, \quad s \equiv \frac{\partial p}{\partial T},
\end{equation}
where $T$ denotes the temperature of the plasma, $s$ is the entropy, and we have $w = e + p$.
For a FOPT system, the energy $e$ and pressure $p$ (the EoS) are
\begin{align}
e(\Phi,T) &= aT^4 + V_{\rm eff}(\Phi,T) - T\frac{\partial V_{\rm eff}}{\partial T},\\
p(\Phi,T) &= \frac{1}{3}aT^4 - V_{\rm eff}(\Phi,T),
\end{align}
where $a = g_*\pi^2/30$ and $g_*$ is the number of degree of freedom, and  $V_{\rm eff}(\Phi, T)$ is the finite temperature effective potential of the scaler fields, which can be calculated with a standard method~\cite{Quiros:1999jp} for a given particle physics model.

Now, the energy-momentum tensor of the whole phase transition system is
\begin{equation}
T^{\mu\nu} = \partial^\mu\phi_i\partial^\nu\phi_i - g^{\mu\nu}\left(\frac{1}{2}\partial_{\alpha}\phi_i\partial^{\alpha}\phi_i\right) + wu^\mu u^\nu - g^{\mu\nu}p,
\end{equation}
where the metric $g^{\mu\nu}=\mathrm{diag}\{+---\}$.
The energy-momentum conservation gives
\begin{equation}
\partial_\mu T^{\mu\nu} = \partial_\mu T_{\phi}^{\mu\nu} + \partial_\mu T_{\rm pl}^{\mu\nu} = 0.\label{eq:EMTc}
\end{equation}
At the bubble wall, we have to fully consider the contribution of the scalar fields and the plasma to the energy-momentum tensor.
When particles penetrate the bubble wall, they should experience an out-of-equilibrium phenomenon, and this should give an effective friction force to the bubble wall.
Here we introduce a phenomenological dissipative (or friction) term, which
parameterizes the out-of-equilibrium phenomenon and also acts as an effective friction. %introduced by the deviation from thermal equilibrium. 
This friction term is also responsible for the entropy production at the bubble wall.
And eq.~\eqref{eq:EMTc} can be further expressed by
\begin{equation}
\partial_\mu T^{\mu\nu} =  \partial_\mu T_{\rm \phi}^{\mu\nu} + \partial_\mu T_{\rm pl}^{\mu\nu} = \chi^\nu - \chi^\nu = 0,
\end{equation}
where we choose $\chi^\nu = \eta u^\mu\partial_\mu\phi_i\partial^\nu\phi_i$ as in refs.~\cite{Ignatius:1993qn,Kurki-Suonio:1995yaf}. 
Note that there are also other choices for this friction term as suggested in refs.~\cite{Sopena:2010zz,Huber:2011aa,Huber:2013kj}.
Substituting $\chi^\nu$ into eq.~\eqref{eq:EMphi} and eq.~\eqref{eq:EMpl}, we have
\begin{align}
\partial_\mu T_{\phi}^{\mu\nu} = \partial_\mu \partial^\mu\phi_i\partial^\nu\phi_i + \frac{\partial V_{\rm eff}}{\partial\phi_i}\partial^\nu\phi_i = -\eta u^\mu\partial_\mu\phi_i\partial^\nu\phi_i,\label{eq:eomphif}\\
\partial_\mu T_{\rm pl}^{\mu\nu} = \partial_\mu(wu^\mu u^\nu) - \partial^\nu p + \frac{\partial V_{\rm eff}}{\partial\phi_i}\partial^\nu\phi_i = \eta u^\mu\partial_\mu\phi_i\partial^\nu\phi_i.\label{eq:eomplf}
\end{align}
The FOPT proceeds with bubble nucleation, expansion, and collision. To gain a complete understanding  of the dynamics of an FOPT system,  numerical hydrodynamical simulations mush be performed, although they could be rather computationally difficult and expensive.
Fortunately, after bubble nucleation, the bubble will quickly reach a steady-state stage and expand with a terminal bubble wall velocity $v_w$.
As a result, for simplicity, we can focus on dynamics of the system during the steady-state stage, where the time-dependence can be neglected in the wall frame.
Then, the system can be divided into three regions: the symmetric phase (the plasma outside the bubble), the broken phase (the plasma in the bubble), and the bubble wall.
In the symmetric and broken phases, the plasma can be treated as being in thermal equilibrium, with the contribution of the scalar fields to the energy momentum tensor $T_{\phi}^{\mu\nu}$ being constant.
Hence, we can just analyze the dynamics of the perfect fluid based on energy-momentum conservation, $\partial_\mu T_{\rm pl}^{\mu\nu} = 0$.
While in the region of the bubble wall, the plasma encounters an out-of-equilibrium phenomenon, which is parameterized by the friction term $\chi^\nu$, and the scalar fields vary violently; and the energy-momentum conservation law of the plasma $\partial_\mu T_{\rm pl}^{\mu\nu} = 0$ fails in this region.
Therefore,  both scalars and plasma must be considered in this region with $\partial_\mu T^{\mu\nu} = \partial_\mu T_{\phi}^{\mu\nu} + \partial_\mu T_{\rm pl}^{\mu\nu} = 0$.
The dynamics of these three regions should be discussed independently, as detailed in the following subsections.

\subsection{Across the bubble wall}
At the bubble wall, we have to fully consider the contribution of the scalar fields and the plasma to the energy-momentum tensor.
From eq.~\eqref{eq:eomphif} and eq.~\eqref{eq:eomplf}, without assuming any symmetries, we have
\begin{align}
\ddot{\phi} - \nabla^2\phi + \frac{\partial V_{\rm eff}}{\partial \phi} &= -\eta \gamma(\dot{\phi} + v^i\partial_i\phi),\\
\dot{E} + \partial_i(Ev^i) + p[\dot{\gamma} + \partial_i(\gamma v^i)] - \frac{\partial V_{\rm eff}}{\partial\phi}\gamma(\dot{\phi} + v^i\partial_i\phi) &= \eta \gamma^2 (\dot{\phi} + v^i\partial_i\phi)^2,\\
\dot{Z}_i + \partial_j(Z_iv^j) + \partial_i p + \frac{\partial V_{\rm eff}}{\partial\phi}\partial_i\phi &= -\eta \gamma(\dot{\phi} + v^j\partial_j\phi)\partial_i\phi.
\end{align}
where $v^i = u^i/\gamma$, $E = \gamma e$ and $Z_i = \gamma^2(e+p)v_i$.
While for simplicity, some symmetries are inserted in these equations, and we just discuss the case with a single scalar\footnote{Generalizing to the case with multiple scalar fields  is straightforward.} henceforth.
In this work, we assume a planar symmetry, which enforces $u^\mu = \gamma(1, 0, 0, v)$, then we have 
\begin{align}
\ddot{\phi} - \partial_z^2\phi + \frac{\partial V_{\rm eff}}{\partial \phi} &= -\eta \gamma(\dot{\phi} + v\partial_z\phi),\\
\partial_\mu (wu^\mu) u^\nu + wu^\mu\partial_\mu u^\nu - \partial^\nu p &= \left[\eta \gamma(\dot{\phi} + v\partial_z\phi) + \frac{\partial V_{\rm eff}}{\partial \phi}\right]\partial^\nu\phi.
\end{align}
At the steady-state stage, the time-derivative can be neglected in the wall frame. 
Therefore, the steady-state profiles across the bubble wall can be derived by the following differential equations
\begin{align}
\partial_z^2\phi - \frac{\partial V_{\rm eff}}{\partial \phi} - \eta \gamma v\partial_z\phi &= 0,\label{eq:phieom}\\
\partial_z\left[w\gamma^2v\right] &= 0,\label{eq:pl1}\\
\partial_z\left[\frac{1}{2}(\partial_z\phi)^2 + w\gamma^2v^2 + p\right] &= 0.\label{eq:pl2}
\end{align}
For different hydrodynamical modes, the initial conditions of these differential equations are different.
Once the initial conditions are fixed, we can derive the spatial dependent scalar field $\phi(z)$, velocity $v(z)$, and temperature $T(z)$ profiles across the bubble wall.
These profiles actually give various quantities ($\phi_+$, $\phi_-$, $v_+$, $v_-$, $T_+$, and $T_-$) just in front of and behind the bubble wall, where quantities in symmetric phase are denoted with subscript $+$, while $-$ represents quantities in the broken phase. These are the key quantities for conducting the model-dependent analysis of the kinetic energy fraction.

\subsection{Away from the bubble wall}
The dynamics of the plasma in the broken and symmetric phases can be descried by the energy-momentum conservation law of the perfect fluid.
Inserting some specific symmetries, we have \cite{Kurki-Suonio:1984zeb,Leitao:2010yw}
\begin{align}
\partial_t[(e + pv^2)\gamma^2] + \partial_r[(e + p)\gamma^2v] &= -\frac{j}{r}[(e + p)\gamma^2v],\\
\partial_t[(e + p)\gamma^2v] + \partial_r[(ev^2 + p)\gamma^2] &= -\frac{j}{r}[(e + p)\gamma^2v^2],
\end{align}
where $j = 2$, $1$, and $0$ for spherical, cylindrical and planar symmetry, respectively.
Note that the covariant derivative and the different metrics should be employed to derive the spherical and cylindrical cases.
$r$ is the distance from the symmetry point, axis and plane, and $t$ is the elapsed time since the bubble nucleation.
At the steady-state stage, both the bubble and fluid profiles are self-similar, which means they maintain their relative shape but rescale as the bubble expands.
This self-similar solution only depends on $\xi = r/t$, thus we have $\partial_t = -(\xi/t)\partial_{\xi}$ and $\partial_r = (1/t)\partial_{\xi}$.
Therefore, we can derive 
\begin{align}
(\xi - v)\frac{\partial_\xi e}{w} &= j\frac{v}{\xi} + [1 - \gamma^2v(\xi - v)]\partial_\xi v,\\
(1 - v\xi)\frac{\partial_\xi p}{w} &= \gamma^2(\xi - v)\partial_{\xi}v,
\end{align}
where $\partial_\xi e$ and $\partial_\xi p$ can be related by the sound speed of the plasma, which is defined as $c_s^2 \equiv dp/de = (dp/dT)/(de/dT)$.
Hence, the sound speed of the plasma is temperature-dependent in general.
Then the equations describing the velocity, enthalpy, and temperature profile are
\begin{align}
j\frac{v}{\xi} &= \gamma^2(1 - v\xi)\left[\frac{\mu^2}{c_s^2(T)} - 1\right]\partial_{\xi}v,\label{eq:vpf}\\
\frac{\partial_{\xi}w}{w} &= \left[1 + \frac{1}{c_s^2(T)}\right]\mu\gamma\partial_{\xi}v,\label{eq:wpf}\\
\frac{\partial_\xi T}{T} &= \gamma^2\mu\partial_\xi v,\label{eq:tpf}
\end{align}
where 
\begin{equation}
\mu(\xi, v) = \frac{\xi - v}{1 - \xi v}.
\end{equation}
With appropriate boundary conditions, we can solve the  velocity $v(\xi)$, enthalpy $w(\xi)$, and temperature $T(\xi)$ profiles.
The boundary conditions can be obtained by the conventional model-independent method~\cite{Espinosa:2010hh,Leitao:2010yw,Leitao:2014pda,Giese:2020rtr,Giese:2020znk,Wang:2020nzm,Wang:2022lyd}, based on various approximations.
In this work, we propose a model-dependent method that preserves the temperature- and position-dependent nature of the sound speed to derive the corresponding boundary conditions.
In the following section, we will go over both methods in detail. 

\section{Kinetic energy fraction}
\label{sec:Kdefdet}

For a thermal cosmological FOPT, the latent heat released by the phase transition must be transferred into the gradient energy of the bubble wall, as well as the thermal energy and kinetic energy of the ambient plasma.
And the so-called "energy budget" of the cosmological FOPT refers the amount of energy obtained by the bubble wall and the plasma.
After bubble collisions, the gradient energy stored in the bubble wall and the kinetic energy of the plasma will generate GW signals.
According to recent studies~\cite{Caprini:2015zlo,Caprini:2019egz}, during the phase transition, GW signals can be produced by three mechanisms, which are bubble collisions, sound waves, and turbulence; and the dominant source of the phase transition GWs is the sound waves in a thermal FOPT.
As a result, the kinetic energy fraction $K$, which denotes the fraction of the total energy of the bubble that is converted to the kinetic energy of the ambient plasma, is a crucial parameter in determining the strength of GW spectrum.
In fact, the GW spectrum from sound wave mechanism scales as $h^2\Omega_{\rm GW} \propto K^2$ or $h^2\Omega_{\rm GW} \propto K^{3/2}$.
According to the energy-momentum tensor of the perfect fluid, i.e., eq.~\eqref{eq:EMpl}, the kinetic energy density of the fluid is \cite{Leitao:2010yw}
\begin{equation}
    e_{\rm kin} = T^{00}(v) - T^{00}(0) = wv^2\gamma^2.
\end{equation}
The volume of the planar, cylindrical, and spherical bubble (i.e. the volume of bubble in one, two, and three space dimension case) can be expressed as
\begin{equation}
    \label{eq:vb}
    V_b = C_j R_b^{j + 1}/(j + 1),
\end{equation}
where $C_j$ is an irrelevant factor in our calculation, and $R_b = v_wt$ is the bubble radius.
The kinetic energy of the fluid is 
\begin{equation}
    \label{eq:E_K}
    E_{\rm kin} = C_j\int wv^2\gamma^2R^jdR.
\end{equation}
%The kinetic energy fraction is defined as the ratio of the kinetic energy to the energy the bubble contain,
Based on the definition of the kinetic energy fraction (i.e. the ratio of the kinetic energy to the total energy the bubble contain), we have
\begin{equation}
    K \equiv \frac{E_{\rm kin}}{e_+V_b}.
\end{equation}
By plugging in eq.~\eqref{eq:vb} and \eqref{eq:E_K}, we have
\begin{equation}
    K = \frac{j + 1}{e_+v_w^{j + 1}}\int wv^2\gamma^2\xi^jd\xi,\label{eq:kfrac}
\end{equation}
then, for the spherical bubble, we have
\begin{equation}
\begin{split}
K_{\rm sph} \equiv \frac{\rho_{\rm fl}^{\rm sph}}{e_+} &= \frac{3}{e_+ v_w^3}\int w(\xi)v^2\gamma^2\xi^2d\xi,\\
\rho_{\rm fl}^{\rm sph} &= \frac{3}{v_w^3}\int w(\xi)v^2\gamma^2\xi^2d\xi,
\end{split}\label{eq:Ksph}
\end{equation}
while for the planar bubble, we have
\begin{equation}
\begin{split}
K_{\rm pla} \equiv \frac{\rho_{\rm fl}^{\rm pla}}{e_+} &= \frac{1}{e_+ v_w}\int w(\xi)v^2\gamma^2d\xi,\\
\rho_{\rm fl}^{\rm pla} &= \frac{1}{v_w}\int w(\xi)v^2\gamma^2d\xi.
\end{split}\label{eq:Kpla}
\end{equation}
To derive the kinetic energy fraction, both the enthalpy $w(\xi)$ and velocity profiles $v(\xi)$ of different hydrodynamical modes must be calculated properly.
Contrary to the conventional model-independent method, we propose a model-dependent method to compute the kinetic energy fraction. 
In the following, we first review possible hydrodynamical modes and the conventional model-independent method then present the model-dependent analysis method in details.

\begin{figure}[ht!]
	\centering
	\includegraphics[width=0.8\textwidth]{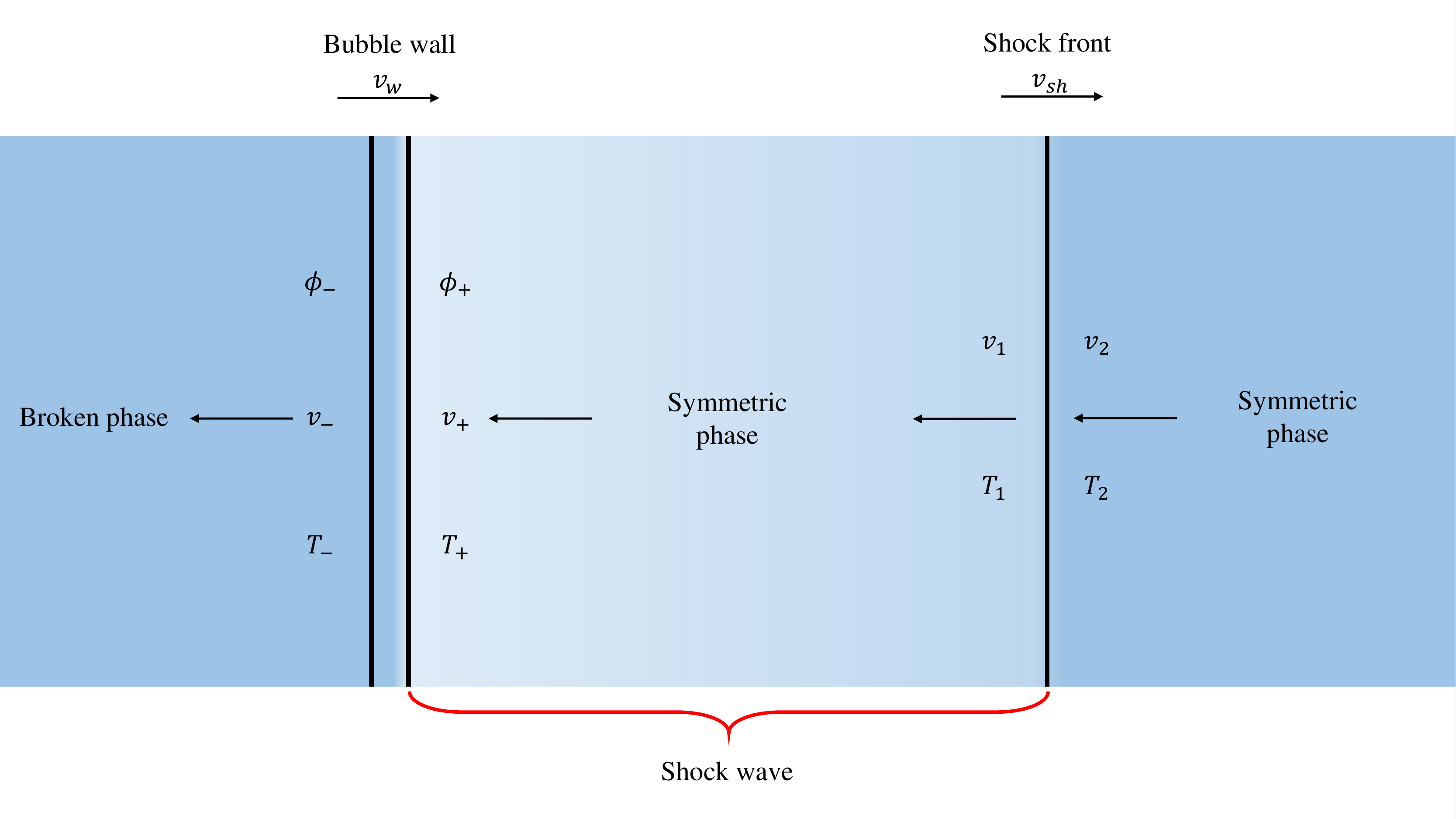}
	\includegraphics[width=0.8\textwidth]{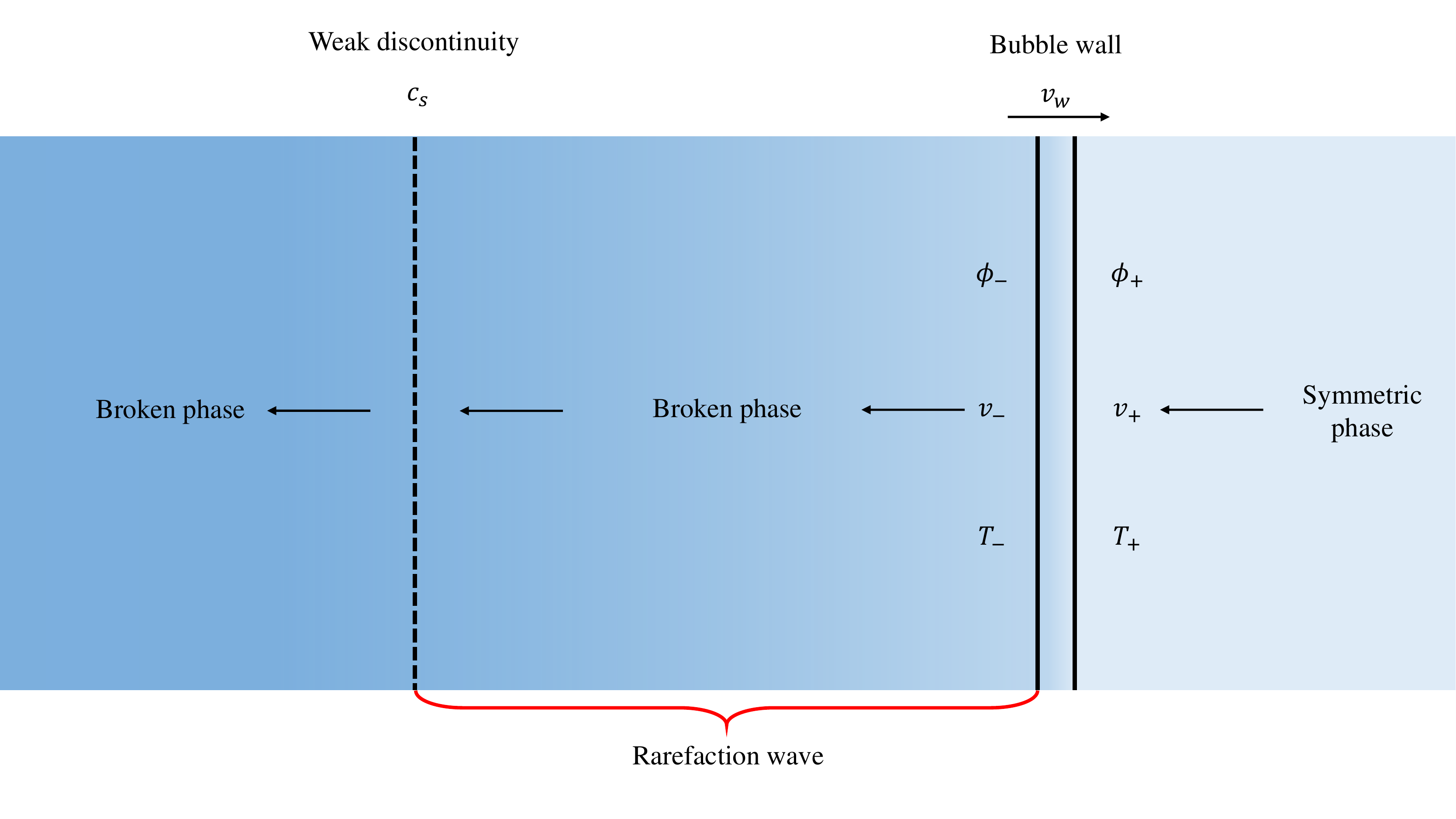}
	\caption{Illustration of field and fluid quantities for detonation and deflgration. The top panel represents a deflagration mode, the bottom panel depicts a detonation mode.}\label{fig:defdet}
\end{figure}

\subsection{Hydrodynamical modes}

There are two kinds of hydrodynamical modes in a thermal FOPT: the detonation mode and the deflagration mode.
These modes are described by physical quantities shown in figure~\ref{fig:defdet}, where the top panel depicts a typical deflagration, while the bottom panel shows a typical detonation.
In the rest frame of the bubble wall, the fluid velocity of the incoming flow is denoted as $v_+$, while the outgoing fluid velocity is $v_-$.
For a detonation mode, the fluid is decelerated by the bubble wall ($v_+ > v_-$), and the incoming flow is supersonic ($v_+ > c_s$).
Whereas for deflagrations, the outgoing fluid is accelerated by the bubble wall ($v_- > v_+$), and $v_+ < c_s$.
While in the rest frame of the shock front, the incoming fluid velocity is represented by $v_2$ and the outgoing fluid velocity is denoted by $v_1$.
Depending on the outgoing flow is subsonic or supersonic, these two modes can be further divided into weak ($v_- < c_s$), Jouguet ($v_- = c_s$), and strong ($v_- > c_s$) detonations, and weak ($v_- < c_s$), Jouguet ($v_- = c_s$), and strong ($v_- > c_s$) deflagrations. 
However, previous studies~\cite{Huet:1992ex,Kurki-Suonio:1995rrv,Kurki-Suonio:1995yaf} have shown that only three modes are stable in an FOPT, namely, weak detonation, weak deflagration, and Jouguet deflagration (i.e., the supersonic deflagration or hybrid).
In the rest part of this work, the terms "detonations" and "deflagrations" refer to weak detonations and weak deflagrations.
And the Jouget deflagration (hybrid) mode is beyond the scope of this work and will be discussed in the future work.

In the rest frame of the bubble center, or equivalently the plasma frame, the fluid velocity is denoted by $\tilde{v}$.
Then the incoming fluid velocity is $\tilde{v}_+$ and the outgoing fluid velocity is $\tilde{v}_-$.
As shown in figure~\ref{fig:defdet}, the detonation forms a rarefaction wave behind the wall, while the deflagration results in a shock wave just in front of the wall.
The fluid must be at rest at the bubble center and far in front of from the bubble wall.
Therefore, in a detonation mode, the fluid in front of the wall is at rest when it is hit by the wall, and the wall is followed by a rarefaction wave, which smoothly reduces the fluid velocity $\tilde{v}$ to zero at $\xi = c_{s,-}$, acting as a weak discontinuity.
In the deflagration mode, however, a shock wave that is able to heat up the fluid proceeds in front of the wall, and the fluid behind the wall is at rest.
In addition, the fluid in front of the shock front is also at rest.
The shock front is a discontinuity that expands with a constant velocity $v_{sh}$.

\subsection{Model-independent analysis}

\begin{figure}[ht]
	\centering
	\includegraphics[width=0.95\textwidth]{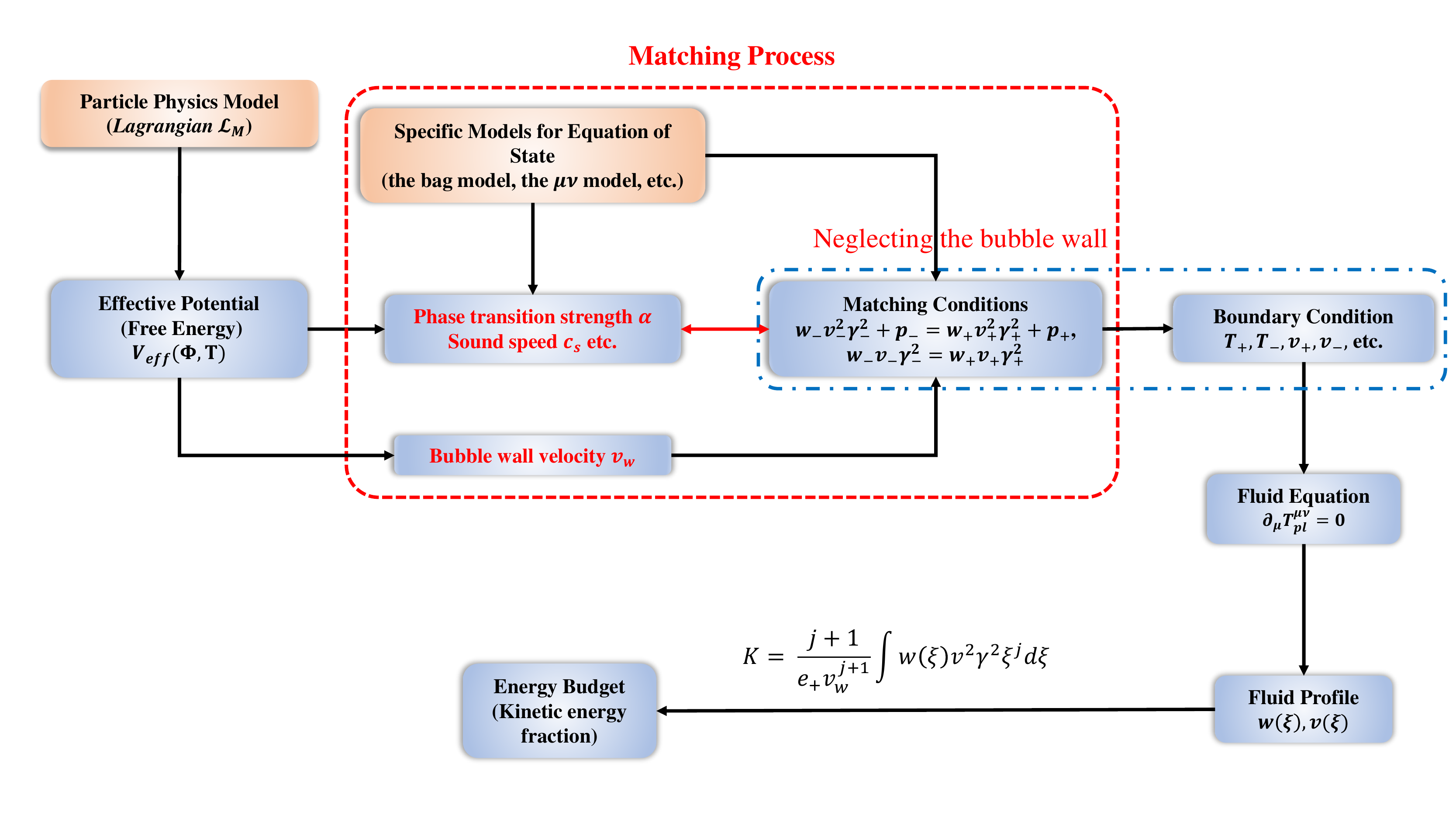}
	\caption{Illustration for model-independent analysis method.}\label{fig:mid}
\end{figure}

Here, we first review the conventional model-independent method. Figure~\ref{fig:mid} demonstrates the whole process of the model-independent analysis with the detail procedure described as follows:
\begin{itemize}
    \item Construct a specific model for the EoS first. This model should be able to approximately represent the broken and symmetric phases as generally as possible.
    \item Derive the matching conditions according to the energy-momentum conservation law by treating the bubble wall as a discontinuity.
    \item Define the parameters that are crucial for matching a particle physics model to the specific model of EoS one constructed, for example, the strength parameter $\alpha$, the sound speed $c_s$, etc.
    \item Calculate the effective potential of some well-motivated particle physics models with a conventional or improved method.
    \item Compute the characteristic temperature $T_*$, the strength parameter $\alpha$, the sound speed $c_s$, and other possible parameters, which depend on the constructed EoS model, based on the effective potential.
    \item Estimate the bubble wall velocity $v_w$ with some appropriate method or manually give it a reasonable value.
    \item Substitute the strength parameter $\alpha$, the bubble wall velocity $v_w$, the sound speed $c_s$, and other possible parameters into the matching condition and derive the boundary conditions.
    \item Solve the fluid equation with boundary conditions of different hydrodynamical modes, then obtain the corresponding fluid profiles.
    \item Relying on the fluid profiles of different hydrodynamical modes, i.e., weak detonation, weak deflagration, and hybrid, finally derive the kinetic energy fraction with eq.~\eqref{eq:kfrac}.
\end{itemize}

Next we describe the model-independent method in more detail.
In most studies, the above procedure is usually performed using the bag model of EoS.
For the bag model, in the symmetric phase, we have
\begin{equation}
p_+ = \frac{1}{3}a_+T_+^4 - \epsilon, \quad e_+ = a_+T_+^4 + \epsilon,\label{eq:eosbagp}
\end{equation}
while in the broken phase, we have
\begin{equation}
p_- = \frac{1}{3}a_-T_-^4, \quad e_- = a_-T_-^4,\label{eq:eosbgm}
\end{equation}
where $\epsilon$ is the bag constant, which characterises the false-vacuum energy.
To observe the logic of using the bag model to describe the FOPT, we can start with the effective potential, which is also the free energy of the whole FOPT system.
The validity of using bag model to describe the FOPT system can be established by analyzing an effective potential.
For a given particle physics model, the effective potential can be divided into two parts
\begin{equation}\label{freeen}
\mathcal{F}(\phi, T) = V_{\rm eff}(\phi, T) = V_{T=0}(\phi) + V_T(\phi,T)\,\,,
\end{equation}
where the one-loop zero temperature part is $V_{T = 0}(\phi)$, while the one-loop thermal correction part $V_T(\phi,T)$ is 
\begin{equation}
V_T(\phi,T) = \pm T\int \frac{d^3 k}{(2\pi)^3} \ln(1 \mp e^{-\omega_k/T})=\pm \frac{T^4}{2\pi^2} J_{b/f}\left(\frac{m}{T}\right)\,\,,
\end{equation}
where the pure temperature dependent part $aT^4/3$ is absorbed in this formula, and $\omega_k^2 = m^2 + k^2 $, the thermal function is represented as
\begin{equation}
J_{b/f}(y^2) = \int_{0}^{\infty}dxx^2\ln\left(1 \mp e^{-\sqrt{x^2+ y^2}}\right)\,\,.
\label{eq:thfunc}
\end{equation}
At the high temperature limit, $m/T\ll1$, the thermal function can be expanded as 
\begin{equation}\label{jb}
\begin{split}
J_b(y^2) \approx &-\frac{\pi^4}{45} + \frac{\pi^2}{12}y^2 - \frac{\pi}{6}(y^2)^{3/2} - \frac{1}{32}y^4\ln\frac{y^2}{a_b}\,\,,\\
%&-2\pi^{7/2}\sum_{l=1}^{\infty}\frac{(-1)^l}{(l+2)!}\Gamma\left(l+\frac{1}{2}\right)\left(\frac{y^2}{4\pi^2}\right)^{l+2}\zeta(2l + 1)\\
a_b = &16\pi^2\exp\left(\frac{3}{2} - 2\gamma_E\right)\,\,,
\end{split}
\end{equation}
for bosons.
While, for fermions, the thermal function can be approximately expanded as
\begin{equation}\label{jf}
\begin{split}
J_f(y^2) \approx &\frac{7\pi^4}{360} - \frac{\pi^2}{24}y^2 - \frac{1}{32}y^4\ln\frac{y^2}{a_f}\,\,,\\
%&-\frac{\pi^{7/2}}{4}\sum_{l=1}^{\infty}(-1)^l\frac{\zeta(2l+1)}{(l+1)!}(1 - 2^{-2l -1})\Gamma\left(l + \frac{1}{2}\right)\left(\frac{y^2}{\pi^2}\right)^{l + 2}\\
a_f =& \pi^2\exp\left(\frac{3}{2} - 2\gamma_E\right)\,\,,
\end{split}
\end{equation}
where $y^2 = m^2/T^2$, $\ln a_b = 5.4076$, $\ln a_f = 2.6351$ and $\gamma_E\approx0.5772$ represents the Euler-Masccheroni constant.
If only the first term of the high temperature expansion of thermal functions is preserved, we can have
\begin{equation}
\mathcal{F}(\phi, T) \approx -\frac{aT^4}{3} + V_{T=0}(\phi),
\end{equation}
and this is exactly the bag model, and the bag constant $\epsilon$ is indeed the zero temperature part $V_{T=0}(\phi)$.
Hence, the bag model gives the same sound speed in both phase, $c_{s,-}^2 = c_{s,+}^2 = 1/3$.
However, keeping only the first term of the high temperature expansion is a rather crude approximation.
Higher order terms of $y$ can definitely give significant modifications.
Taking the $\mathcal{O}(y^2)$ terms into account, the free energy is
\begin{equation}
\mathcal{F}(\phi, T) \approx -\frac{aT^4}{3} + bT^2 + V_{T=0}(\phi),
\label{eq:f2o}
\end{equation}
where $b = \sum_ig_i\tilde{c_i}m_i^2(\phi)/24$, and $i$ runs over the particles that acquire a masses during an FOPT, $g_i$ is the number of degree of freedom, $\tilde{c}_i = 1/2$ for fermions and $1$ for bosons.
Apparently, according to the definition of sound speed ($c_s^2 = dp/de$ and $-\mathcal{F} = p$), the $\mathcal{O}(y^2)$ will modify it.
To include the corrections to the sound speed from higher order terms, one construct the $\mu\nu$ model~\cite{Leitao:2014pda,Giese:2020rtr,Giese:2020znk,Wang:2020nzm}, which takes the sound speed of both phases as extra parameters, and the sound speed is still constant in this model. 
However, from eq~\eqref{eq:f2o}, we can find the sound speed is actually temperature dependent, and the temperature around the bubble wall is position-dependent, which makes the sound speed position-dependent.
Recently, ref~\cite{Wang:2022lyd} constructed a model of EoS to describe the temperature-dependent sound speed by introducing more parameters.
Nevertheless, the accuracy is only kept at specify orders of $y$.
To resolve this problem, we propose a model-dependent method in the next subsection.

With a given model of EoS, the matching conditions can be written as
\begin{equation}
w_-v_-^2\gamma_-^2 + p_- = w_+v_+^2\gamma_+^2 + p_+,\quad w_-v_-\gamma_-^2 = w_+v_+\gamma_+^2\,\,,\label{eq:mcond}
\end{equation}
which are obtained from the energy-momentum conservation $\partial_\mu T_{\rm pl}^{\mu\nu} = 0$ across the wall.
Note that the bubble wall is treated as a discontinuity, and the contribution to the energy-momentum tensor from the bubble wall is neglected.
Then the strength parameter in the bag model to carry out the matching process can be defined as 
\begin{equation}
    \alpha = \frac{\epsilon}{a_+T_+^4}.
\end{equation}
Note that the definition of strength parameter is usually more complicated in other more sophisticated models of EoS~\cite{Leitao:2014pda,Giese:2020rtr,Giese:2020znk,Wang:2020nzm,Wang:2022lyd}.
From the effective potential, we can calculate the bubble wall velocity $v_w$, strength parameters $\alpha$, sound speed $c_s$, and other possible parameters at characteristic temperature $T_*$.
Then the matching process can be done after obtaining all those quantities, and we can derive the boundary condition for the fluid equations~\eqref{eq:vpf},~\eqref{eq:wpf}, and \eqref{eq:tpf}.
Based on the boundary conditions, the fluid profiles for different hydrodynamical modes can be derived, which finally give the kinetic energy fraction.

\subsection{Model-dependent analysis}

\begin{figure}[ht]
	\centering
	\includegraphics[width=0.95\textwidth]{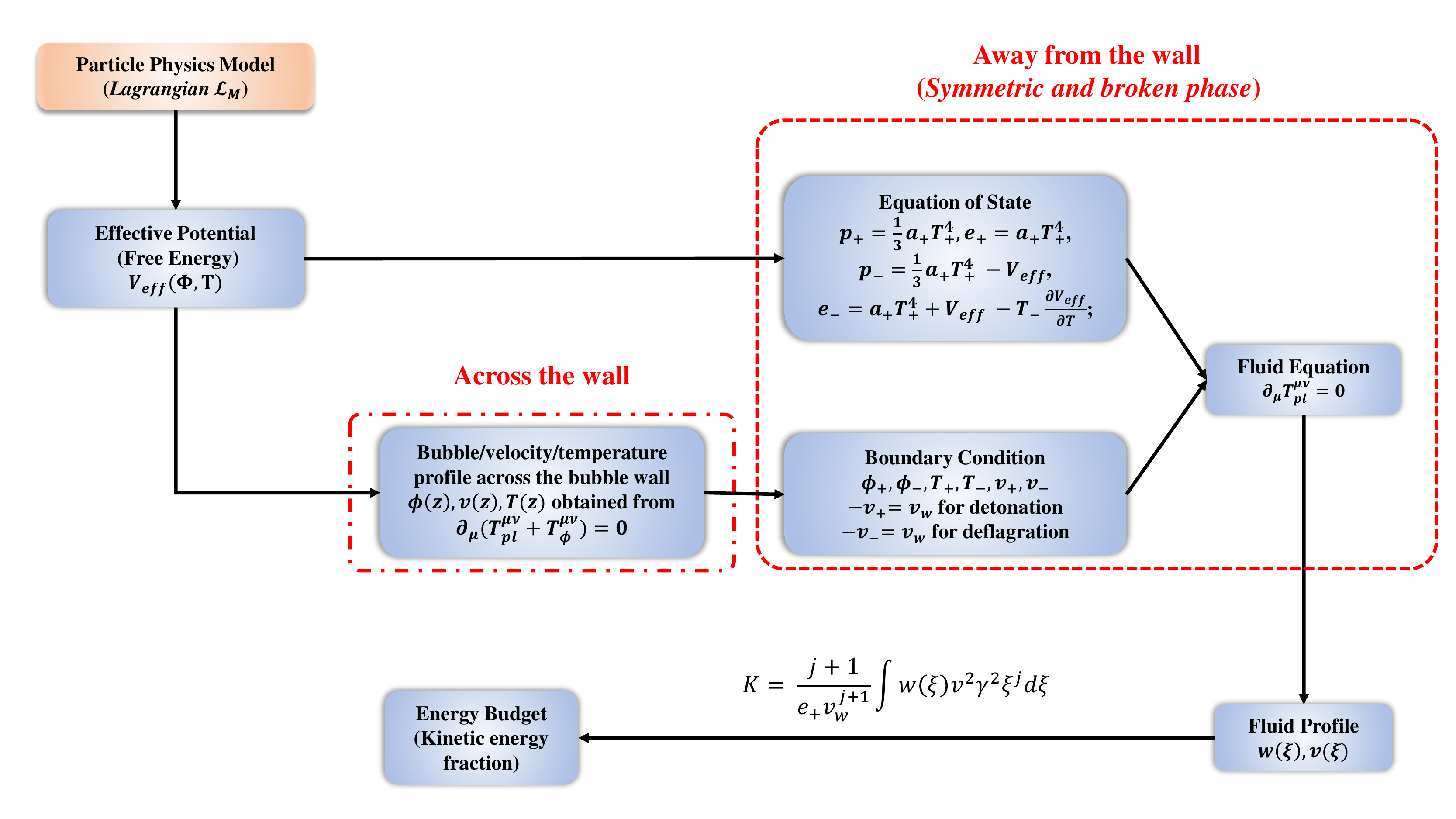}
	\caption{Illustration for model-dependent analysis method.}
	\label{fig:mdm}
\end{figure}

Through the matching process, it is straightforward to quantify the energy budget for different particle physics models using the model-independent method.
However, the accuracy of this method depends on the model of EoS, and we also need to separately calculate the bubble wall velocity.
Furthermore, recent studies~\cite{Balaji:2020yrx,Ai:2021kak,Laurent:2022jrs,Wang:2022txy} have shown that the hydrodynamics of both phases has a rather strong effect on the bubble wall velocity.
Hence, it is more reasonable to combine the calculation of the hydrodynamics and the bubble wall velocity.
Since the effective potential of the particle physics model is the free energy of the phase transition system, we can actually derive the EoS directly from the effective potential without constructing any additional EoS.
Therefore, we propose a model-dependent analysis method for calculating the energy budget, and the basic idea of this method is shown in figure~\ref{fig:mdm}.
In the following, we demonstrate the whole procedure:
\begin{itemize}
    \item Calculate the effective potential for a given specific particle physics model.
    
    \item Derive the characteristic temperature with the standard method.
    
    \item Solve the local equations of motion across the bubble wall.
    Here, for simplicity, we adopt the planar approximation, the local EoM across the bubble wall can be further simplified as
    \begin{align}
    \partial_z^2\phi - \frac{\partial V_{\rm eff}}{\partial \phi} - \tilde{\eta} T_c \gamma v\partial_z\phi &= 0,\label{eq:pphi}\\
    w\gamma^2v &= C_1,\label{eq:pc1}\\
    \frac{1}{2}(\partial_z\phi)^2 + w\gamma^2v^2 + p &= C_2,\label{eq:pc2}
    \end{align}
    where $C_1$ and $C_2$ are constants, and we introduce a dimensionless parameter $\tilde{\eta}$, where $\tilde{\eta}T_c = \eta$, and $T_c$ is the critical temperature.
    
    \item Obtain the velocity $v(z)$, the temperature $T(z)$, and the scalar field $\phi(z)$ profile across the bubble wall, which give the boundary conditions of the fluid equations.
    
    \item Derive the EoS for both phases. 
    According to the effective potential, the EoS of symmetric phase can be written as
    \begin{equation}
    p_+ = \frac{1}{3}a_+T_+^4, \quad e_+ = a_+T_+^4,\label{eq:eosp}
    \end{equation}
    where we implicitly assume $\Phi_+ = 0$\footnote{However, if we consider a multi-step phase transition, this is not necessarily true.},
    while in the broken phase, the EoS is
    \begin{equation}
    \begin{split}
    e_- &= a_-T_-^4 + V_{\rm eff}(\Phi_-,T_-) - T_-\frac{\partial V_{\rm eff}}{\partial T},\\
    p_- &= \frac{1}{3}a_-T_-^4 - V_{\rm eff}(\Phi_-,T_-),
    %w &= \frac{4}{3}aT^4 - cT^2\phi^2 + ET\phi^3
    \label{eq:eosm}
    \end{split}
    \end{equation}
    
    \item Substitute the eq.~\eqref{eq:eosp} and eq.~\eqref{eq:eosm} into the fluid equations~\eqref{eq:vpf}, \eqref{eq:wpf}, and \eqref{eq:tpf}, with the boundary conditions given by the profiles across the bubble wall, the fluid profiles for different hydrodynamical modes can be solved.
    
    \item Using the fluid profiles of different hydrodynamical modes, the kinetic energy fraction can be eventually obtained with eq.~\eqref{eq:kfrac}.
\end{itemize}
In the following two subsections, for different hydrodynamical modes, we describe how to calculate the local EoM across the bubble wall and derive the corresponding profiles in detail.

\subsubsection{Deflagration}
Deflagration modes develop a shock front in the symmetric phase, which is a discontinuity.
From the energy-momentum conservation across the shock front $\partial_{\mu}T_{\rm pl}^{\mu\nu} = 0$, we can obtain the following relations of relevant quantities in the shock front frame: 
\begin{equation}
v_1v_2 = c_{s,+}^2, \quad \frac{v_1}{v_2} = \frac{T_2^4 + c_{s,+}^2T_1^4}{T_1^4 + c_{s,+}^2T_2^4},\label{eq:shmc}
\end{equation}
where $c_{s,+}$ is the sound speed of the symmetric phase, and $c_{s,+} = 1/\sqrt{3}$ for an one-step phase transition.
Therefore, we have
\begin{equation}
v_1 = \frac{1}{\sqrt{3}}\sqrt{\frac{3T_2^4 + T_1^4}{3T_1^4 + T_2^4}},\quad v_2 = \frac{1}{3v_1},\label{eq:sh_v12}
\end{equation}
and 
\begin{equation}
T_2 = T_*, \quad v_2 = v_{sh}.
\end{equation}
In the equations above, all of the velocities are represented by their absolute values.
In figure~\ref{fig:defdet}, the directions of bubble wall and shock front velocities are to the right, while the fluid velocities are to the left.
Hence, if the bubble wall and shock front velocities are positive, the fluid velocity is then negative.
For the deflagration mode, the procedures for computing the bubble and fluid profiles across the wall are shown in the following:
\begin{enumerate}
	\item Guess the bubble wall velocity $v_w$ ($v_w < c_s$ for deflagration mode) and the temperature just behind the shock front $T_1$, and derive the constants $C_1$ and $C_2$ at $z = z_{\rm sh}$ ($z_{\rm sh}$ is the position of the shock front) with the following relations
	\begin{equation}
	\frac{4}{3}aT_1^4\frac{v_{1, \rm wf}}{1 - v_{1,\rm wf}^2} = C_1, \quad \frac{4}{3}aT_1^4\frac{v_{1, \rm wf}^2}{1 - v_{1, \rm wf}^2} + \frac{1}{3}aT_1^4 = C_2,
	\end{equation}
	where $v_{1,\rm wf}^2$ is the fluid velocity just behind the shock front in the wall frame.
	And according to eqs.~\eqref{eq:shmc} and \eqref{eq:sh_v12}, the absolute value of $v_{1,\rm wf}$ can be calculated through the following formula: 
	\begin{align}
	\tilde{v}_1 &= \frac{v_2 - v_1}{1 - v_2v_1} = \frac{1 - 3v_1^2}{2v_1},\\
	v_{1, \rm wf} &= \frac{v_w - \tilde{v}_1}{1 - v_w\tilde{v}_1},
	\end{align}
	where $\tilde{v}_1$ is the fluid velocity just behind the shock front in the plasma frame.
	
	\item Derive the initial conditions of $T$ and $\phi$ at $z = 0$ from the following equations
	\begin{align}
	\left[w_r-T\frac{\partial V_{\rm eff}}{\partial T}(\phi_0^{\rm def},T_0^{\rm def})\right]\gamma_w^2v_w &= C_1,\\
	\left[w_r-T\frac{\partial V_{\rm eff}}{\partial T}(\phi_0^{\rm def},T_0^{\rm def})\right]\gamma_w^2v_w^2 + p(\phi_0^{\rm def},T_0^{\rm def}) &= C_2,
	\end{align}
	and the initial conditions of deflagration are
	\begin{equation}
	\phi_0^{\rm def} = f^{\rm def}(v_w, T_1),\quad
	\phi_0^{'\rm def} = 0,\quad
	T_0^{\rm def} = g^{\rm def}(v_w, T_1),\quad
	v_0^{\rm def} = v_w,
	\end{equation}
	where $f^{\rm def}$ and $g^{\rm def}$ are specific functions of $T_1$ and $v_w$.
	
	\item Solve the differential equations \eqref{eq:pphi} and calculate $v$ and $T$ at the same time. 
	If we have $v(+\infty)=v_{1, \rm wf}$,  $\phi(+\infty) = 0$ and $T(+\infty) = T_1$, the problem is solved. 
	Otherwise, we repeat the above steps until the condition is fulfilled.
	\textit{Note: if we can not find a bubble wall velocity below $1/\sqrt{3}$ that satisfy the conditions, then the deflagration solution dose not exist for a given friction parameter $\tilde{\eta}$.}
\end{enumerate}

\subsubsection{Detonation}

For a detonation mode, the fluid is at rest in front of the bubble wall in the plasma frame.
Hence, we have
\begin{equation}
v_+ =  v_w, \quad T_+ = T_*.
\end{equation}
The procedure of calculating field and fluid profiles across the bubble wall is shown in the following:
\begin{enumerate}
	\item Guess the bubble wall velocity $v_w$ and the fluid velocity just behind the wall $v_0$, then derive $C_1$ and $C_2$ at $z = +\infty$ with the subsequent relations
	\begin{equation}
	\frac{4}{3}aT_+^4\frac{v_+}{1 - v_+^2} = C_1, \quad \frac{4}{3}aT_+^4\frac{v_+^2}{1 - v_+^2} + \frac{1}{3}aT_+^4 = C_2.
	\end{equation}
	
	\item Derive the initial conditions of $T$ and $\phi$ at $z = 0$ from the following equations
	\begin{align}
	\left[w_r(T_0^{\rm det})-T\frac{\partial V_{\rm eff}}{\partial T}(\phi_0^{\rm det},T_0^{\rm det})\right]\gamma_w^2v_w &= C_1,\\
	\left[w_r(T_0^{\rm det})-T\frac{\partial V_{\rm eff}}{\partial T}(\phi_0^{\rm det},T_0^{\rm det})\right]\gamma_w^2v_w^2 + p(\phi_0^{\rm det},T_0^{\rm det}) &= C_2.
	\end{align}
	Then the initial conditions of the detonation mode are
	\begin{equation}
	\phi_0^{\rm det} = f^{\rm det}(v_w,v_0),\quad
	\phi_0^{'\rm det} = 0,\quad
	T_0^{\rm det} = g^{\rm det}(v_w, v_0),\quad
	v_0^{\rm det} = v_0,
	\end{equation}
	where $f^{\rm det}$ and $g^{\rm det}$ are specific functions of $v_0$ and $v_w$.
	
	\item Solve the differential equations \eqref{eq:pphi} and compute $v$ and $T$ at the same time. 
	If we have $T(+\infty)=T_*$, $\phi(+\infty) = 0$ and $v(+\infty) = v_w$, the problem is solved. 
	%If the solution can not match the condition at $z = +\infty$, 
	Otherwise, we repeat the above steps until the condition is fulfilled.
	\textit{Note: if we can not find a bubble wall velocity satisfies the conditions, then the detonation solution dose not exist for a given friction parameter $\tilde{\eta}$.}
\end{enumerate}
In this work, the friction parameter $\tilde{\eta}$ is treated as a free parameters for simplicity.
However, in principle, this parameter is calculable for a given particle physics model~\cite{Huber:2013kj,Megevand:2012rt,Megevand:2013hwa,Konstandin:2014zta}.

\section{A representative model}
\label{sec:model}

In the following, we apply our model-dependent method to a representative model for illustration.
The representative model is
\begin{equation}\label{eq:veff}
V_{\rm eff}(\phi,T)\approx \frac{1}{2}(m^2 + cT^2)\phi^2 - ET \phi^3 + \frac{\lambda}{4}\phi^4 \,\,.
\end{equation}
where $m^2 = - \lambda v_0^2$, and vacuum expectation value $v_0=246.22 ~\rm GeV$ for electroweak phase transition.
This representative model can denote a large class of particle physics models~\cite{Carena:1996wj,Espinosa:1993bs,Huang:2017kzu,Huang:2017rzf,Wang:2019pet,Jiang:2022btc}, such as the standard model, the two-Higgs doublet model, the minimal supersymmetry model, the singlet model, and the hidden phase transition model, etc.
In this model, the lowest temperature $T_0$ where the symmetric minimum still exists is
\begin{equation}
T_0^2 = \frac{\lambda v^2}{c}.
\end{equation}
The critical temperature where the degenerate minimum exists is
\begin{equation}
T_c^2 = \frac{T_0^2}{1 - 2E^2/(\lambda c)}.
\end{equation}
When the temperature is below the critical temperature, the global minimum is
\begin{equation}
\phi_{\rm true} = \frac{3ET + \sqrt{9E^2T^2 - 4\lambda(m^2 + cT^2)}}{2\lambda}.
\end{equation}
Note that when temperature is slightly higher than the critical temperature, this point becomes a local minimum.
According to eq.~\eqref{eq:veff}, we can further derive the corresponding EoS of this model as
\begin{equation}
\begin{split}
p &= \frac{1}{3}aT^4 - \frac{m^2 + cT^2}{2}\phi^2 + ET \phi^3 - \frac{\lambda}{4}\phi^4,\\
e &= aT^4 + \frac{m^2 - cT^2}{2}\phi^2 + \frac{\lambda}{4}\phi^4,
\end{split}\label{eq:eostoy}
\end{equation}
then the enthalpy can be obtained as
\begin{equation}
w = \frac{4}{3}aT^4 - cT^2\phi^2 + ET\phi^3,
\end{equation}
and the temperature-dependent sound speed is
\begin{equation}
c_s^2(T) = \frac{1}{T}\frac{(-aT^4/3 + V_{\rm eff})'}{(-aT^4/3 + V_{\rm eff})''},\label{eq:cs}
\end{equation}
where prime represents the total derivative with respect to temperature $T$.
We assume that the bubble achieves steady-sate stage immediately after nucleation. 
Then the temperature far in front of the bubble wall can be approximated as nucleation temperature\footnote{Refs.~\cite{Ellis:2018mja,Caprini:2019egz,Wang:2020jrd,Athron:2022mmm} suggest the percolation temperature should be a better choice for the characteristic temperature in the calculation of phase transition GW.} $T_n$, and the nucleation temperature can be calculated using the standard method.
Here, we give the model parameters and the corresponding phase transition parameters of the representative model in table~\ref{tb:bpoint1}, where we have defined $\mathcal{S}_*\equiv T_*/T_c$ to quantify the amount of supercooling at the characteristic temperature $T_*$. 
Hence, $\mathcal{S}_n$ and $\mathcal{S}_0$ represent the amount of supercooling at $T_n$ and $T_0$, respectively.
$\alpha_n$ denotes the phase transition strength defined by the bag model at $T_n$. 
To further compare our result with the results given by the $\mu\nu$ model~\cite{Giese:2020rtr,Giese:2020znk,Wang:2020nzm}, in which EoS are given the form 
\begin{align}
    &p_+ = \frac{1}{3}a_+T^\mu - \epsilon, \quad e_+ = \frac{1}{3}a_+(\mu - 1)T^\mu + \epsilon,\label{eq:munup}\\
    &p_- = \frac{1}{3}a_-T^\nu, \quad e_- = \frac{1}{3}a_-(\nu - 1)T^\nu,\label{eq:munum}
\end{align}
where $\mu = 1 + 1/c_{s,+}^2$, $\nu = 1 + 1/c_{s,-}^2$, we also include the corresponding strength parameter $\alpha_{\bar{\theta},*}$, which are defined by
\begin{equation}
    \alpha_{\bar{\theta},*} =  \frac{\Delta\bar{\theta}}{3w_+}, \quad \bar{\theta} = e - p/c_{s,-}^2,
\end{equation}
and the sound speed of the broke phase $c_{s,-}^2$ in table~\ref{tb:bpoint1}.

{\it Note: The benchmark parameters chosen for this representative model are just for illustration purpose without considering any experimental constraints.}
\begin{table}[!t]\small%
	\centering
	\footnotesize% fontsize
	\setlength{\tabcolsep}{2.5pt}% column separation
	\renewcommand{\arraystretch}{1.2}%row space 
	\begin{tabular}{|cccccccccccccc|}
		\hline
		 &$c$ & $E$ & $\lambda$ & $a$ & $T_n ~\rm [GeV]$ & $T_c~\rm [GeV]$ &  $\phi_n~\rm [GeV]$ & $T_0 ~\rm [GeV]$ & $\alpha_n$ & $\alpha_{\bar{\theta},n}$ & $c_{s,-}^2$ & $\mathcal{S}_n$ & $\mathcal{S}_0$\\
		\hline
		BP1&$1.5$ & $0.05$ & $0.1$ & $35.119$ & $64.508592$ & $64.660540$ &  $71.719753$ & $63.573731$ & $0.0128$ & $0.0129$ & $0.1718$ & $0.9977$ & $0.9832$\\
		\hline
		BP2&$1.5$ & $0.07$ & $0.1$ & $35.119$ & $65.373815$ & $65.758110$ &  $103.714539$ & $63.573731$ & $0.0254$ & $0.0258$ & $0.1793$ & $0.9942$ & $0.9668$\\
		\hline

  BP3&$1.5$ & $0.125$ & $0.1$ & $35.119$ & $69.283562$ & $71.450705$ &  $204.047449$ & $63.573731$ & $0.0780$ & $0.0816$ & $0.2092$ & $0.9697$ & $0.8898$\\
		\hline
	\end{tabular}
	\caption{The benchmark parameter sets and the corresponding phase transition parameters of the representative model.
		Here $T_n$ is the nucleation temperature, $T_c$ is the critical temperature, $T_0$ is the lowest temperature where the symmetric metastable minimum $\phi = 0$ still exits, and $a = 106.75 \pi^2/30$.
}\label{tb:bpoint1}
\end{table}

\begin{table}[!t]
	\centering
	\footnotesize% fontsize
	\setlength{\tabcolsep}{2.5pt}% column separation
	\renewcommand{\arraystretch}{1.2}%row space 
	\begin{tabular}{|cccccccc|}
		\hline
		&$\phi_-\rm~[GeV]$ & $T_-\rm~[GeV]$ & $v_-$ & $\phi_+\rm~[GeV]$ & $T_+\rm~[GeV]$ & $v_+$ & $\tilde{\eta}$\\
		\hline
		BP1& 67.120739& 64.613934  & -0.1 & 0 & 64.636932  & -0.096591& 0.198779\\
		\hline
		BP2&98.043840 & 65.581548 & -0.1 & 0 &  65.634895 & -0.093161 &0.566406\\
		\hline
		BP3& 199.965752& 69.724364 & -0.1 & 0 &  70.165663& -0.078261 &2.542273\\
		\hline
	\end{tabular}
	\caption{The boundary conditions of deflagration modes. Note: The fluid velocity is given in the wall frame, and $v_- = -v_w$.}
	\label{tb:bpdf_fric}
\end{table}

\begin{table}[!t]
	\centering
	\footnotesize% fontsize
	\setlength{\tabcolsep}{2.5pt}% column separation
	\renewcommand{\arraystretch}{1.2}%row space 
	\begin{tabular}{|cccccccc|}
		\hline
		&$\phi_-\rm~[GeV]$ & $T_-\rm~[GeV]$ & $v_-$ & $\phi_+\rm~[GeV]$ & $T_+\rm~[GeV]$ & $v_+$ & $\tilde{\eta}$\\
		\hline
		BP3&175.161138 & 71.650425& -0.871573&0 & 64.305635& -0.9&0.273980\\
		\hline
	\end{tabular}
	\caption{Boundary conditions of a detonation mode. Note: the fluid velocity is given in the wall frame, and $v_+ = -v_w$.}
	\label{tb:bp3det_fric}
\end{table}

\subsection{Profiles across the wall}

\begin{figure}[!t]
    \centering
    \includegraphics[width=\textwidth]{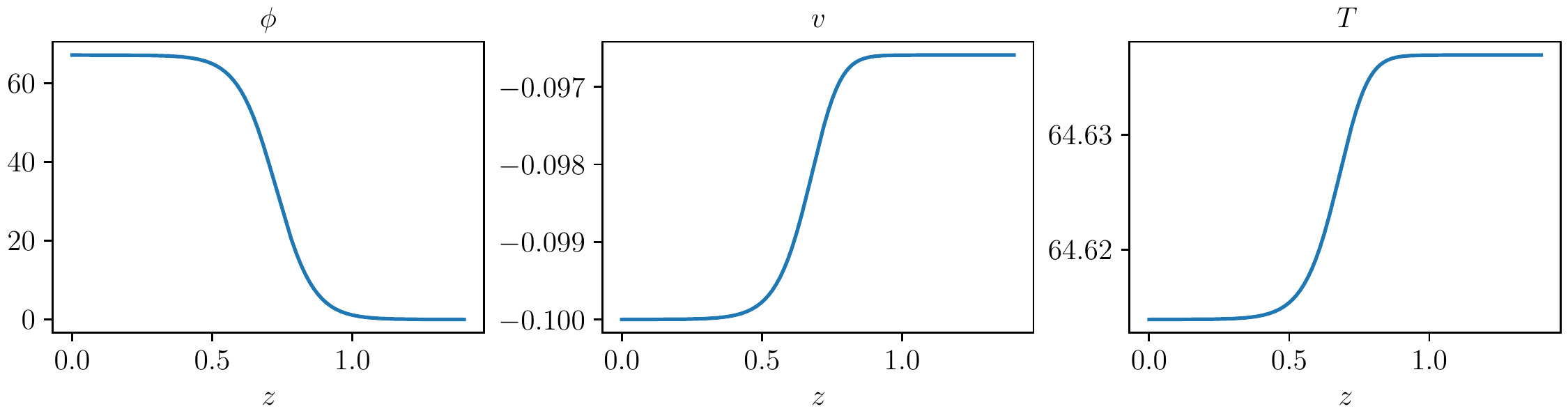}
    \includegraphics[width=\textwidth]{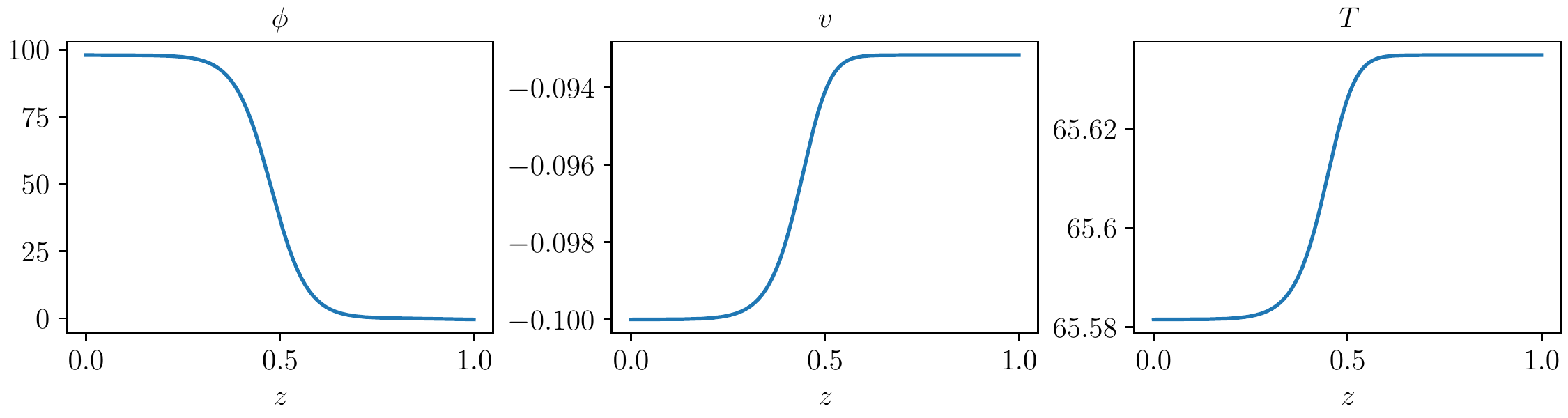}
    \includegraphics[width=\textwidth]{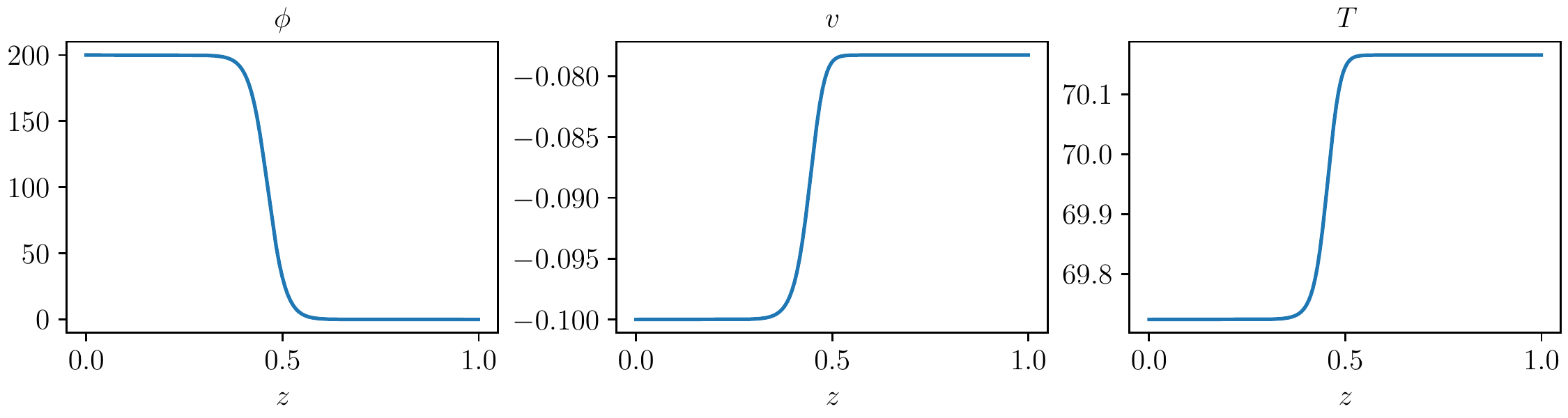}
    \caption{The bubble, velocity and temperature profiles
    of BP1 (top panel), BP2 (middle panel), and BP3 (bottom panel)  for the deflagration mode. The parameter sets are given in table.~\ref{tb:bpoint1}.
    We fix the bubble wall velocity as $v_w = -v_- = 0.1$, and the corresponding friction parameters are given in table~\ref{tb:bpdf_fric}.
    The units of $\phi$ and $T$ are $[\mathrm{GeV}]$, and the units of $z$ are $[\mathrm{GeV}^{-1}]$.}
    \label{fig:bpsdef}
\end{figure}

\begin{figure}[!t]
    \centering
    \includegraphics[width=\textwidth]{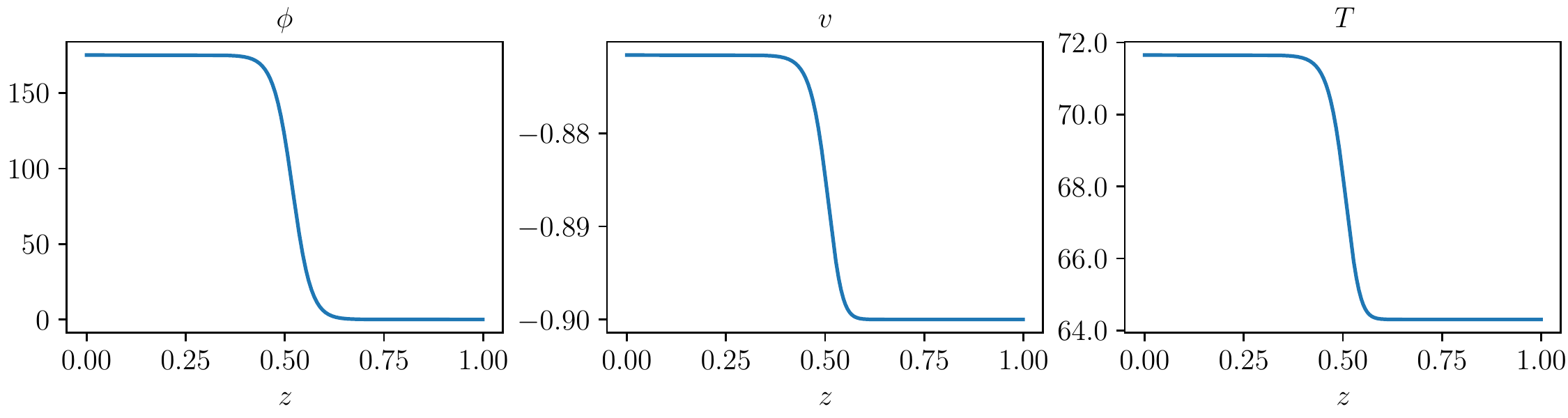}
    \caption{The bubble, velocity and temperature profiles
    of BP3 for the detonation mode. The parameter sets are given in table.~\ref{tb:bpoint1}. We fixed the bubble wall velocity as $v_w = -v_+ = 0.9$ and $T_+ = 0.9~T_c$. The corresponding friction parameters are given in table~\ref{tb:bp3det_fric}. The units of $\phi$ and $T$ are $[\mathrm{GeV}]$, and the units of $z$ are $[\mathrm{GeV}^{-1}]$.}
    \label{fig:bp3det}
\end{figure}

According to the benchmark sets given in table~\ref{tb:bpoint1}, we can calculate the bubble, velocity, and temperature profiles across the bubble wall from eqs.~\eqref{eq:pphi},~\eqref{eq:pc1}, and \eqref{eq:pc2} with the model-dependent method we mentioned in section~\ref{sec:Kdefdet}.
However, to derive a consistent deflagration or detonation solution, we demand a further constraint, which is
\begin{equation}
    \frac{\partial V_{\rm eff}}{\partial \phi} (\phi_-, T_-) = \frac{\partial V_{\rm eff}}{\partial \phi} (\phi_+, T_+) = 0.
\end{equation}
For the deflagration mode, we fixed the bubble wall velocity as $v_w = -v_+ = 0.1$ for simplicity; the friction parameter $\tilde{\eta}$ is treated as a free parameter, and the corresponding friction parameters of these benchmark points are given in table~\ref{tb:bpdf_fric}.
However, in a realistic model, the friction parameters should be calculated first, and the bubble wall velocity can be derived after that.
Here, in figure~\ref{fig:bpsdef}, for the deflagration mode, we show the bubble $\phi(z)$, velocity $v(z)$ and temperature $T(z)$ profiles of three parameter sets.
The top panel shows the corresponding profiles across the bubble wall of BP1, the middle panel denotes the profiles of BP2, and the bottom panel depicts the profiles of BP3.
To derive a consistent detonation solution, we need to vary the value of the friction parameter $\tilde{\eta}$.
But in the frictionless case, where $\tilde{\eta} = 0$, a detonation mode can not be found for all three benchmark points when $T_+ = T_n$, where $T_+$ denotes the temperature just in front of the bubble wall.
As a result, we have to increase the amount of supercooling by decreasing $T_+$ ($T_0 < T_+ < T_n$).
Then, a consistent detonation mode can be constructed for BP3 when $T_+ = 0.9~T_c$.
According to our study and refs~\cite{Ignatius:1993qn,Huber:2013kj}, we can conclude that a detonation mode can only exist when the supercooling is large enough at nucleation temperature, which results from a small $\mathcal{S}_*$ or, equivalently, a large enough $\alpha$.
And this actually requires an effective potential having a deep global minimum and relatively low potential barrier between the global and local minimum at characteristic temperature.
Figure~\ref{fig:bp3det} shows the detonation profiles of BP3 when $T_+ = 0.9~T_c$, and the corresponding friction parameters are given in table~\ref{tb:bp3det_fric}.
After we derive profiles across the bubble wall, the boundary conditions of the fluid eqs.~\eqref{eq:vpf}, \eqref{eq:wpf}, and \eqref{eq:tpf} can be simultaneously derived.
In table~\ref{tb:bpdf_fric}, we provide the boundary conditions of the deflagration mode for all three benchmark points.
We find the temperature of the fluid behind the bubble wall $T_-$ is larger than the temperature far in front of the bubble wall for each benchmark point. 
Therefore, the fluid behind the bubble wall is also heated for a deflagration mode.
For the detonation mode, the corresponding boundary conditions are given in table~\ref{tb:bp3det_fric}.
With these boundary conditions, we can derive the fluid profiles in both phases for different hydrodynamical modes.

\subsection{Fluid profiles}

With the boundary conditions given in table~\ref{tb:bpdf_fric} and table~\ref{tb:bp3det_fric}, we can solve the fluid equations~\eqref{eq:vpf},~\eqref{eq:wpf}, and \eqref{eq:tpf}, then derive the corresponding fluid profiles for deflagration and detonation modes.
However, to derive the fluid profiles in both phases,  the EoS of both phases are needed.
From eq.~\eqref{eq:eostoy}, the EoS of the representative model in the symmetric phase is
\begin{equation}
p_+(T) = \frac{1}{3}aT^4, \quad e_+(T) = aT^4,\label{eq:toyeosp}
\end{equation}
and the sound speed of the symmetric phase $c_{s,+} = 1/\sqrt{3}$ is a constant.
In the broken phase, we set $\phi_{\rm ture} = \phi_{\rm true}(T_-)$ for simplicity, then the EoS are
%In the broken phase, the EoS are
\begin{equation}
\begin{split}
p_-(T) &= \frac{1}{3}aT^4 - \frac{m^2 + cT^2}{2}\phi_{\rm true}^2 + ET \phi_{\rm true}^3 - \frac{\lambda}{4}\phi_{\rm true}^4,\\
e_-(T) &= aT^4 + \frac{m^2 - cT^2}{2}\phi_{\rm true}^2 + \frac{\lambda}{4}\phi_{\rm true}^4,
\end{split}\label{eq:toyeosm}
\end{equation}
and the corresponding sound speed
\begin{equation}
c_{s,-}^2(T) = \frac{4aT^3 - cT\phi_{\rm true}^2 + E\phi_{\rm true}^3}{12aT^3 - 3cT\phi_{\rm true}^2},\label{eq:csm}
\end{equation}
and it is temperature-dependent.
Note that we assume $a$ has the same value in both phases.
Since we assume the bubble preserves a planar symmetry, for consistency, the planer fluid profiles of this representative model should be constructed.
For the planar approximation, some simple solutions~\cite{Kurki-Suonio:1984zeb,Leitao:2010yw,Leitao:2014pda} as following can thus be derived from the fluid equations:
\begin{align}
\tilde{v}(\xi) = \mathrm{constant},\label{eq:vconst}\\
\tilde{v}_{\rm rar} = \frac{\xi - c_s}{1 - c_s\xi}.\label{eq:vrar}
\end{align}
Next, we show in detail how to construct self-similar planar fluid profiles for deflagration and detonation modes.

\begin{figure}[t!]
    \centering
    \includegraphics[width=\textwidth]{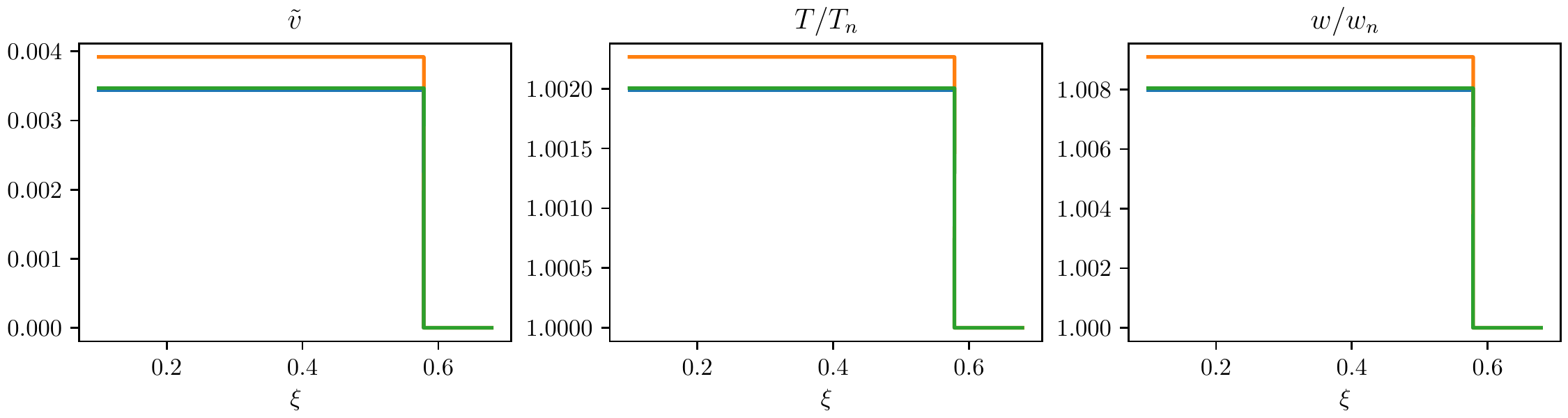}
    \includegraphics[width=\textwidth]{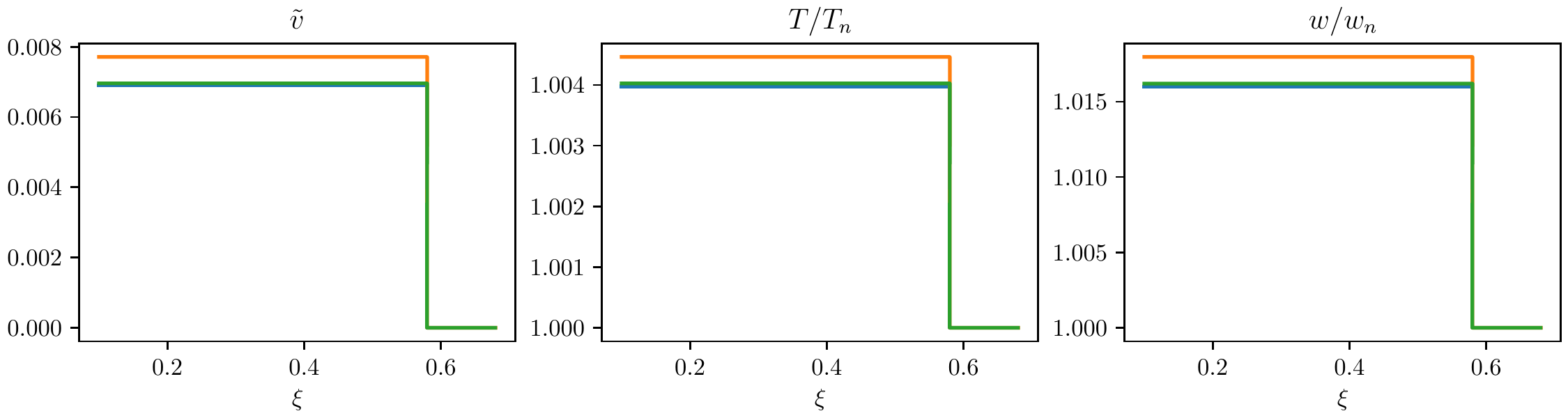}
    \includegraphics[width=\textwidth]{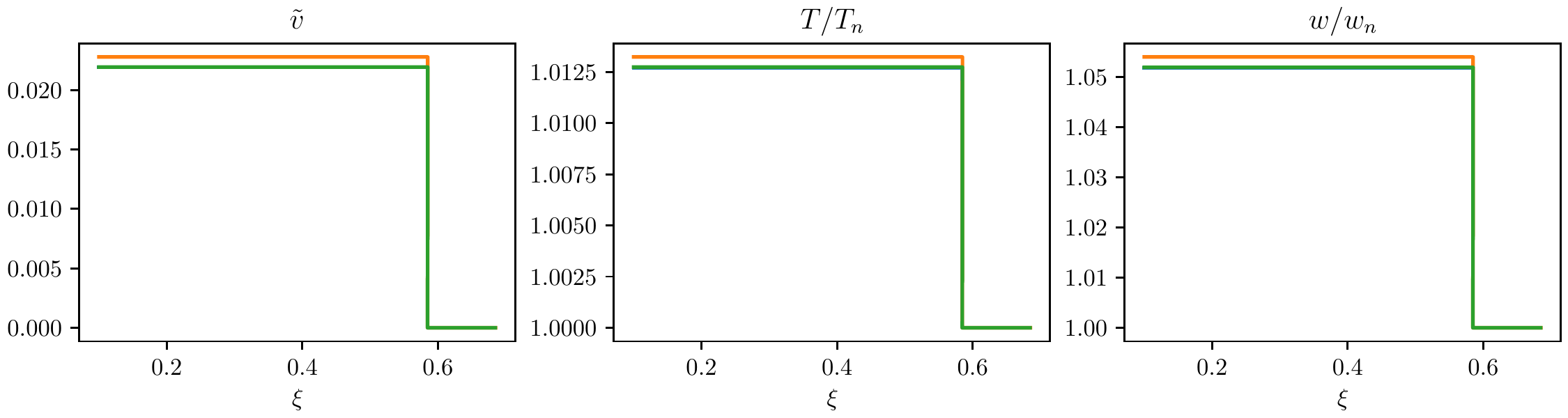}
    \caption{The fluid profiles of BP1 (top panel), BP2 (middle panel), and BP3 (bottom panel) for the deflagration mode. The blue lines denote the results obtained from our model-dependent method, the orange and green lines indicate the profiles derived by model-independent method based on the bag model and the $\mu\nu$ model, respectively. And the green lines almost overlap the blue lines.}
    \label{fig:defpfle}
\end{figure}

\begin{figure}[!t]
    \centering
    \includegraphics[width=\textwidth]{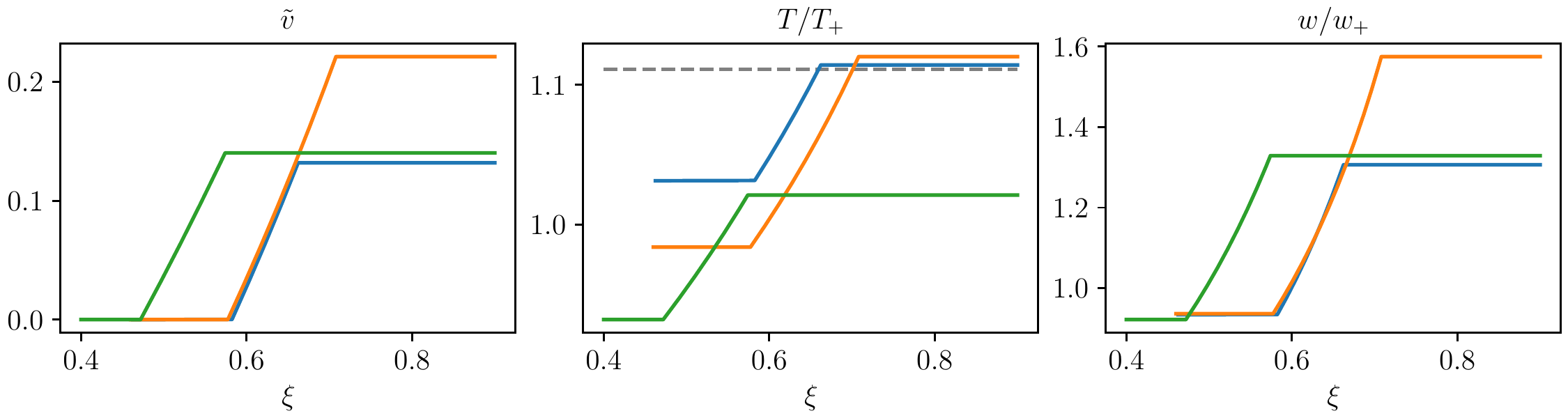}
    \caption{The fluid profiles of BP3 for the detonation mode. The blue lines denote the results obtained from our model-dependent method; the orange and green lines indicate the profiles derived by model-independent method with the bag model and the $\mu\nu$ model, respectively. The grey dashed line in the middle panel denotes the critical temperature.}
    \label{fig:bp3_det}
\end{figure}

For the deflagration mode, in the plasma frame, the turbulent flow is in the symmetric phase and the fluid is at rest in the broken phase.
Since we assume a planar symmetry for the fluid profiles,
the velocity, temperature, and enthalpy profiles should be constant until the shock front based on previous studies~\cite{Kurki-Suonio:1984zeb,Kurki-Suonio:1995rrv,Huet:1992ex,Leitao:2010yw,Leitao:2014pda}.
The position of the shock front can be derived with $\mu(\tilde{v},\xi_{\rm sh})\xi_{\rm sh} = c_{s,+}$, where $c_{s,+} = 1/\sqrt{3}$, and $\tilde{v}$ is the fluid velocity in the plasma frame.
Since, we have already obtained the boundary conditions of fluid velocity in the wall frame and the bubble wall velocity from the local EoM across the bubble wall,
we can derive the fluid velocity in the plasma frame as
\begin{equation}
    \tilde{v}_{\rm def} = \frac{v_w - |v_+|}{1 - v_w|v_+|}.
\end{equation}
With the planar symmetry, fluid profiles of the deflagration mode should be constant, we can derive the corresponding profiles after determining the position of the shock front.
For the representative model, the position of shock front is
\begin{equation}
    \xi_{\rm sh} = \frac{2}{\sqrt{4\tilde{v}_{\rm def}^2 + 12} - 2\tilde{v}_{\rm def}},
\end{equation}
and $\xi_{\rm sh} = v_{\rm sh}$.
At the steady-state stage, the width of bubble wall is negligible compared to the bubble radius~\cite{Huet:1992ex,Kurki-Suonio:1995rrv,Kurki-Suonio:1995yaf,Hindmarsh:2015qta}.
Hence, for the simplicity, the width of the bubble wall is considered as infinitesimally small, and the initial position of the fluid profiles can be set at $\xi_0 = v_w = 0.1$.
Therefore, the corresponding fluid profiles including  velocity $\tilde{v}(\xi)$, temperature $T(\xi)/T_n$, and enthalpy $w(\xi)/w_n$ of BP1 to BP3 are given in figure~\ref{fig:defpfle}.
The top panel denotes the velocity $\tilde{v}(\xi)$, temperature $T(\xi)/T_n$, and enthalpy $w(\xi)/w_n$ profiles of BP1.
The middle panel represents the fluid profiles of BP2, and the bottom panel depicts the fluid profiles of BP3.
The bule lines in figure~\ref{fig:defpfle} are the results derived from the model-dependent method, while the orange and green lines are profiles obtained using the conventional model-independent method with the bag model of EoS and the $\mu\nu$ model of EoS, respectively.
As shown in figure~\ref{fig:defpfle}, the differences of the profiles derived by both methods decrease as the strength of the phase transition grows.
For the deflagration, our method does not show significant differences compared to the model-independent method, and the results derived by the $\mu\nu$ model are almost the same as the model-dependent method.
Since the EoS of the symmetric phase of the representative model is similar to the bag model, both of them only contain the first-order terms of the high-temperature expansion.
However, if we consider a multiple-step phase transition, which can be realized in a particle physics model with multiple scalars, the difference between our method and the conventional method should be more significant.
We will leave the study of the multiple-step phase transition to our future study.

\begin{table}[t]
	\centering
	\footnotesize% fontsize
	\setlength{\tabcolsep}{2.5pt}% column separation
	\renewcommand{\arraystretch}{1.2}%row space 
	\begin{tabular}{|c|ccc|c|}
		\hline
            &\multicolumn{3}{c|}{Deflagration}&Detonation\\
            \hline
		& BP1 & BP2 & BP3 & $\mathrm{BP3}$ \\
		\hline
		$K_{\rm pla}$ & $7.621\times10^{-5}$ & $3.097\times10^{-4}$ & $3.265\times10^{-3}$ & $6.843\times10^{-3}$\\
		\hline
		$K_{\rm pla}^{\rm bag}$ & $9.891\times10^{-5}$ & $3.867\times10^{-4}$ & $3.536\times10^{-3}$ & 
		$2.728\times10^{-2}$\\
		\hline
            $K_{\rm pla}^{\mu\nu}$ & $7.742\times10^{-5}$ & $3.147\times10^{-4}$ & $3.272\times10^{-3}$ & $1.404\times10^{-2}$\\
            \hline
		$\mathcal{R}_{\rm bag}$ & $+29.789\%$ & $+24.875\%$ & $+8.311\%$ & $+298.625\%$\\
		\hline
            $\mathcal{R}_{\mu\nu}$ & $+1.588\%$ & $+1.614\%$ & $+0.214\%$ & $+105.173\%$\\
            \hline
		%$K_s$ & & & &\\
		%\hline
		%$K_s^{\rm bag}$ &  &  & &\\
		%\hline
	\end{tabular}
	\caption{Kinetic energy fraction of the benchmark points for deflagration and detonation. $K_{\rm pla}$, $K_{\rm pla}^{\rm bag}$ and $\mathcal{R}_{\mu\nu}$ represent the kinetic energy fraction derived by the model-dependent method and the conventional model-independent method with the bag model of EoS and the $\mu\nu$ model of EoS, respectively. $\mathcal{R}_{\rm bag} = |K_{\rm pla} - K_{\rm pla}^{\rm bag}|/K_{\rm pla}$ denotes the relative corrections of model-independent method based on the bag model compared to the model-dependent method.
    $\mathcal{R}_{\mu\nu} = |K_{\rm pla} - K_{\rm pla}^{\mu\nu}|/K_{\rm pla}$ is the relative corrections of model-independent method based on the $\mu\nu$ model compared to the model-dependent method.}
	\label{tb:kfraction}
\end{table}

For detonation cases, the fluid profile should be the superposition of eq.~\eqref{eq:vconst} and eq.~\eqref{eq:vrar}.
Namely, the fluid velocity is a constant just behind the wall, then at a specific point the profile should be connected with a rarefaction solution.
The position of the turning point is
\begin{equation}
    \xi_{\rm tp} = \frac{\tilde{v}_- + c_{s,-}}{1 + \tilde{v}_-c_{s,-}}
\end{equation}
where $\tilde{v}_-$ and $c_{s,-}$, which describe the fluid velocity and the sound speed just behind the bubble wall, can be derived from the boundary conditions given in table~\ref{tb:bp3det_fric} and eq.~\eqref{eq:cs}.
In the bag model of EoS or the $\mu\nu$ model, the sound speed $c_s$ is a constant in the broken phase.
However, in more general cases, $c_s = c_s(\xi)$, the sound speed is actually position-dependent.
Substituting eq.~\eqref{eq:vrar} into eq.~\eqref{eq:tpf}, we have the following ordinary differential equations
\begin{equation}
\begin{split}
\partial_{\xi}T &= T\gamma^2 \mu \partial_\xi \tilde{v}_{\rm rar}, \\
\partial_{\xi}\tilde{v}_{\rm rar} &= \frac{d}{d\xi}\left[\frac{\xi - c_s}{1 - c_s\xi}\right]
\end{split}
\end{equation}
Since the initial conditions of the fluid velocity and temperature are obtained from the profiles across the wall,
we can directly derive the rarefaction part of the detonation profile.
With the constant and the rarefaction parts determined, combining them enables us to quantify the fluid profiles of the detonation mode.
In figure~\ref{fig:bp3_det}, we show the fluid profiles of BP3 for a detonation mode with our model-dependent method (blue lines) and compare them to the traditional methods (orange and green lines).
As shown in figure~\ref{fig:bp3_det}, the differences of the profiles derived by different methods are significant for the detonation mode.
Here, we neglect the thickness of the bubble wall based on refs~\cite{Huet:1992ex,Kurki-Suonio:1995rrv,Kurki-Suonio:1995yaf,Hindmarsh:2015qta}, the initial point of the fluid profiles of detonation mode is at $\xi = 0.9$.

With the self-similar fluid profiles shown in figure~\ref{fig:defpfle} and figure~\ref{fig:bp3_det}, we can finally calculate the kinetic energy fraction of this representative model using eq.~\eqref{eq:kfrac}.
In table~\ref{tb:kfraction}, we present the corresponding kinetic energy fractions of the three benchmark points BP1, BP2, and BP3 calculated by the model-dependent and the model-independent methods, respectively.
Here, we use $K_{\rm pla}$ to represent the kinetic energy fraction derived by the model-dependent method. 
While $K_{\rm pla}^{\rm bag}$ and $K_{\rm pla}^{\mu\nu}$ denote kinetic energy fraction obtained by the conventional model-independent method based on the bag EoS and the $\mu\nu$ EoS, respectively.
We also give the relative corrections of model-independent method compared to the model-dependent method.
For the three benchmark points, we observe that in the deflagration mode, the relative corrections range from 10\% to 30\% for the bag model. As for the $\mu\nu$ model, the relative corrections are around 1\%.
This conventional model-independent method based on the $\mu\nu$ model significantly improves the results.
However, for BP3 in the detonation mode, significant differences between the model-dependent and the model-independent methods are found.
The relative corrections for the bag model are close to $300\%$, while for the $\mu\nu$ model, the relative corrections are approximately $100\%$.

\section{Discussion}
\label{sec:dic}

\subsection*{Comparison with the model-independent method}
For the representative model eq.~\eqref{eq:veff}, we derive the fluid profiles across and away from the bubble wall for different hydrodynamic modes in the last section. 
The corresponding kinetic energy fractions derived by our model-dependent method and the conventional model-independent method are also shown in table~\ref{tb:kfraction}.
We compare and discuss the results as follows:
\begin{itemize}
    \item {\it Deflagration mode.} 
    According to the fluid profiles shown in figure~\ref{fig:defpfle} and the kinetic energy fraction given in table~\ref{tb:kfraction}, we find our method yields results that differ from the conventional model-independent method.
    The difference could be attributed to two reasons: the EoS and the treatment of the bubble wall.
    For the EoS, in the symmetric phase, we have eq.~\eqref{eq:toyeosp} for our model-dependent method, whereas the bag model gives the EoS as eq.~\eqref{eq:eosbagp} and the $\mu\nu$ model gives the EoS as eq.~\eqref{eq:munup}. 
    Compared with the bag model and the $\mu\nu$ model, the constant $\epsilon$ is omitted in our method.
    Our EoS of the representative model, the bag model and the $\mu\nu$ model give a constant sound speed $c_{s,+} = 1/\sqrt{3}$.
    Hence the EoS of the symmetric phase are basically the same for the bag model, the $\mu\nu$ model and our representative model.
    To derive the boundary conditions of the deflagration mode, we need to deal with the shock front, which is treated as a discontinuity in both methods. For the bubble wall,  
    the conventional model-independent method treats the bubble wall as a discontinuity, and the matching conditions are obtained by neglecting the bubble wall. This is due to comparing to the bubble radius, the bubble wall thickness is negligible, its contribution should be strongly suppressed.  
    While our model-dependent method need to solve the EoM of the scalar field to derive the corresponding boundary conditions.
    To find the main source of the discrepancies between our method and the conventional model independent methods, we compare the results from those different methods.% based on some representative models.
    
    For the deflagration mode of BP1, the boundary conditions of the fluid velocity and temperature derived by our model-dependent method are $\tilde{v}_+ = 0.0034$ and $T_+/T_n = 1.0020$ respectively, while the boundary conditions derived by the model-independent method based on the bag model are $\tilde{v}_+^{\rm bag} = 0.0039$ and $T_+^{\rm bag}/T_n = 1.0023$, corresponding to the relative corrections of $14.706\%$ and $0.03\%$, respectively.
    And the boundary conditions obtained by the model-independent method based on the $\mu\nu$ model are $\tilde{v}_+^{\mu\nu} = 0.0035$ and $T_+^{\mu\nu}/T_n = 1.00201$, corresponding to the relative corrections of $2.941\%$ and $0.000998\%$, respectively.
    For BP2, the boundary conditions derived by the model-dependent method are $\tilde{v}_+ = 0.0069$ and $T_+/T_n = 1.0040$, while the boundary conditions derived by the model-independent method are $\tilde{v}_+^{\rm bag} = 0.0077$ and $T_+^{\rm bag}/T_n = 1.0045$, corresponding to relative corrections of $11.594\%$ and $0.05\%$, respectively.
    And the boundary conditions obtained by the model-independent method based on the $\mu\nu$ model are $\tilde{v}_+^{\mu\nu} = 0.00695$ and $T_+^{\mu\nu}/T_n = 1.00402$, corresponding to the relative corrections of $0.725\%$ and $0.00199\%$, respectively.
    For BP3, the boundary conditions derived by the model-dependent method are $\tilde{v}_+ = 0.0219$ and $T_+/T_n = 1.0127$, while the boundary conditions derived by the model-independent method are $\tilde{v}_+^{\rm bag} = 0.0228$ and $T_+^{\rm bag}/T_n = 1.0132$, corresponding to relative corrections of $4.12\%$ and $0.049\%$, respectively.
    And the boundary conditions obtained by the model-independent method based on the $\mu\nu$ model are $\tilde{v}_+^{\mu\nu} = 0.02193$ and $T_+^{\mu\nu}/T_n = 1.01274$, corresponding to the relative corrections of $0.137\%$ and $0.039\%$, respectively.
    As we can see, the relative corrections of $\tilde{v}_+$ decrease as the strength of phase transitions increases, whereas the relative corrections of $T_+/T_n$ are small and less affected by the strength of phase transitions.
    However, as the strength of phase transition increases, the absolute differences between the boundary conditions of the velocity and the temperature become larger.

    We find the conventional method based on the $\mu\nu$ model almost gives the same boundary conditions as our model-dependent method. When compared with the bag model, the difference arises from the different EoS of the broken phase used in the calculation.
    Therefore, we can conclude that the boundary conditions are actually strongly affected by the EoS of the broken phase, and the discrepancy between our method and the conventional model-independent method are predominantly originated from different EoS.

    \item {\it Detonation mode.}
    For the detonation mode,  
    we find that the differences of the boundary conditions and kinetic energy fraction are more significant according to figure~\ref{fig:bp3_det} and table~\ref{tb:kfraction}.
    The discrepancies in the detonation mode might be also originated from the EoS and the treatment of the bubble wall.
    The bag model of EoS in the broken phase is given by eq.~\eqref{eq:eosbgm}, and the $\mu\nu$ model gives eq.~\eqref{eq:munum}, while the EoS derived directly by the effective potential of the representative is shown in eq.~\eqref{eq:toyeosm}, which are obviously different.
    This differences can be observed from the thermal function eq.~\eqref{eq:thfunc}. The bag model is equivalent to the leading order approximation of high-temperature expansion of the thermal function. And the $\mu\nu$ incorporated the higher order effect in a constant sound speed that deviates from $1\sqrt{3}$. However, the EoS derived directly from the effective potential preserves the high order corrections and is also capable of capturing the temperature dependency of the sound speed.
    To confirm the dominant source of discrepancies, we also perform the following comparison.
    
    %For the detonation mode, our model-dependent calculation also takes the contribution of the scalar field to energy-momentum tensor across the bubble wall into account.
    %Therefore, both the EoS and the treatment of the bubble wall should be responsible for the evident effect on boundary conditions.
    From our tests, the corresponding fluid profiles and kinetic energy fractions derived by the model-dependent method and the model-independent method confirm the significant differences quantitatively.
    For BP3, we set $T_+ = 0.9~T_c$ to derive the detonation solution, and the corresponding strength parameter at this temperature is $0.14$ for the bag model and 0.15 for the $\mu\nu$ model.
    And the boundary conditions of the fluid velocity and temperature derived by our model-dependent method are $\tilde{v}_- = 0.1319$ and $T_-/T_+ = 1.114$, while we have $\tilde{v}_-^{\rm bag} = 0.2213$ and $T_-^{\rm bag}/T_+ = 1.1202$ for the model-independent method based on the bag model, corresponding to 
    relative corrections of $67.779\%$ and $0.557\%$, respectively.
    For the model-independent method based on the $\mu\nu$ model, $\tilde{v}_-^{\mu\nu} = 0.1402$ and $T_-^{\mu\nu}/T_+ = 1.0737$, corresponding to the relative corrections of $6.293\%$ and $3.618\%$.

    In summary, for the detonation mode, we also find the conventional method based on the $\mu\nu$ model can yield boundary conditions that are close to our results within a few percent.
    When comparing it to the bag model, we conclude that the modification mainly comes from the different EoS of the broken phase.
    Consequently, the discrepancies are predominantly originated from different EoS, and the contribution of the bubble wall is negligible.
    
\end{itemize}

To summarize, our mode-dependent method is based on the effective potential of a specific particle physics model, and the EoS of the phase transition system is directly obtained from the effective potential.
This approach is capable of capturing the temperature dependency of the sound speed of the plasma.
Besides, the bubble wall is not simply treated as a discontinuity, and the contribution of the scalar field to the energy-momentum tensor across the bubble wall is taken into account, though the contribution of the bubble wall is strongly suppressed and negligible.
Meanwhile, since we need to solve the EoM of the scaler field in our method, the bubble wall velocity and the boundary conditions of the fluid equations can be derived simultaneously.
In contrast, the conventional model-independent method relies on matching the particle physics model to a specific model of EoS, usually failing to consider the temperature dependency of the sound speed.
In addition, the EoS model employed in the model-independent method may neglect the effect higher order corrections. Hence, it should be less accurate compared to our model-dependent analysis. 
Indeed, the difference could be significant in some cases.
For example, compared with our model-dependent method, for the kinetic fraction, the relative corrections of the model-independent method based on the bag mode are around 300\% and $10\%-30\%$  for the detonation and the deflagration mode, respectively. 
Since the EoS in the symmetric phase is similar to the bag model, the difference is less significant in the deflagration mode.
Our mode-dependent calculations also show that both the detonation and deflagration modes heat the fluid deep inside bubbles, whereas the conventional model-independent method always shows that the temperature deep inside bubbles is lower than the temperature far in front of the bubble wall.
Moreover, the number of degrees of freedom should be different in the symmetric phase and the broken phase, since particles obtain mass in the broken phase.
For the conventional model-independent method, we can change the value of $a$ in the bag model to incorporate this effect,
while in our model-dependent method, an additional differential equation~\cite{Ignatius:1993qn,Kurki-Suonio:1995rrv,Kurki-Suonio:1995yaf} $T \partial_\mu(su^\mu) = \eta(u^\mu\partial_\mu\phi)^2$ should be considered to take this effect into account.
Further study of this topic is planned as future work.
In addition, to reduce the theoretical uncertainties introduced by different forms of the friction terms in eq.~\eqref{eq:eomphif}, we can actually substitute the Boltzmann equation into the calculation of profiles across the wall and the bubble wall velocity, as shown in refs~\cite{Moore:1995ua,Moore:1995si,John:2000zq,Kozaczuk:2015owa,Laurent:2020gpg,Friedlander:2020tnq,Wang:2020zlf,Dorsch:2021ubz,Dorsch:2021nje,Laurent:2022jrs,Jiang:2022btc}.
And the corresponding improvement of our model-dependent method is in progress.

\subsection*{Effects of bubble geometry}
\begin{figure}[!t]
    \centering
    \includegraphics[width=\textwidth]{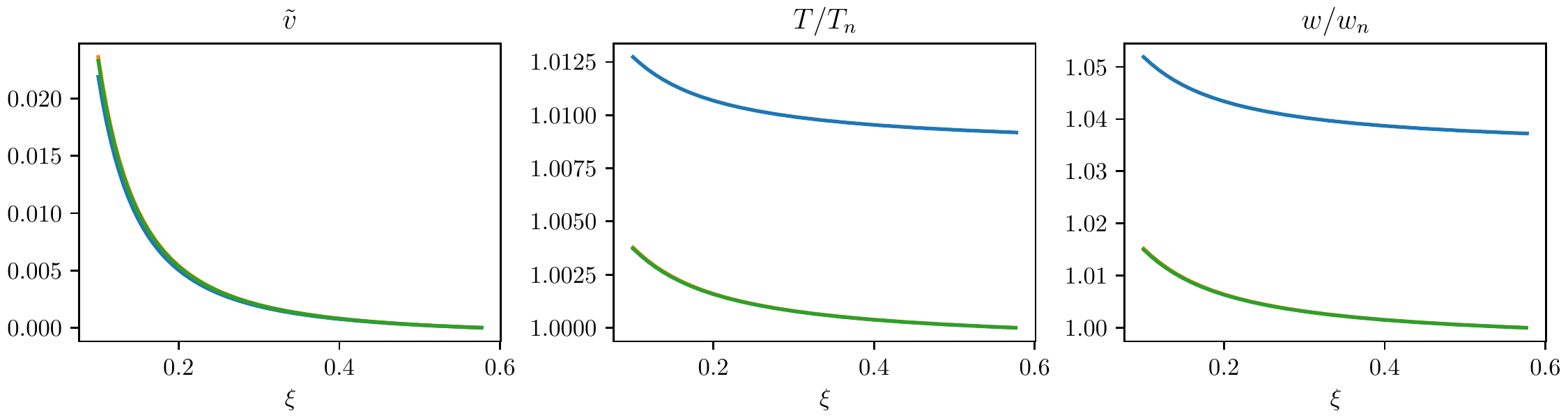}
    \includegraphics[width=\textwidth]{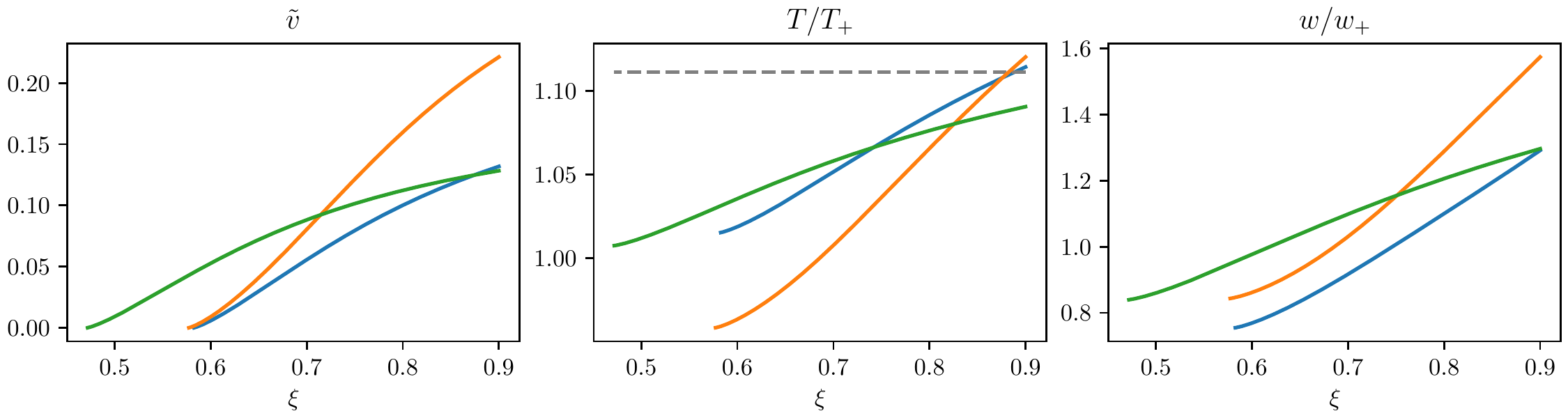}
    \caption{The spherical deflagration (top panel) and detonation (bottom panel) fluid profiles of BP3. The blue lines denote the results obtained from our model-dependent method, the orange and green lines indicate the profiles derived by model-independent method with the bag EoS and the $\mu\nu$ EoS. The grey dashed line denotes the critical temperature.}
    \label{fig:sph}
\end{figure}
To simplify the calculation, we often insert some specific symmetries for the bubbles~\cite{Kurki-Suonio:1984zeb,Leitao:2010yw}, e.g., planar, cylindrical, and spherical symmetries.
In this work, we assume that the local EoM across the bubble wall has a planar symmetry. We then derive the planer fluid profiles with boundary conditions given by the local EoM across the bubble wall.
However, different geometries should give different fluid profiles, and ref.~\cite{Leitao:2010yw} shows that the geometry can give modifications to the kinetic energy fraction of the FOPT. 
Based on the bag model of EoS, ref.~\cite{Leitao:2010yw} studies the effect of bubble geometry using the conventional model-independent method, and they find that the planar case always make the value of kinetic energy fraction larger than the spherical case for the deflagration mode.
However, for the detonation mode, the differences between the planar case and the spherical case are negligibly small.
Since the thickness of the fluid shell is comparable to the bubble radius, the planar approximation of the fluid shell may not be appropriate.
As suggested in ref.~\cite{Sopena:2010zz,Huber:2011aa,Huber:2013kj}, one may use the planer bubble wall to give boundary conditions of the fluid profile, while keeping the sphericity of the fluid profiles.
Following this way,  one can get spherical fluid profiles, as shown in figure~\ref{fig:sph}.
A comparison of the fluid profiles derived by our model-dependent method and the conventional model-independent method based on the bag model and the $\mu\nu$ model is also given in this figure. 
These results can serve as an approximation for the spherical fluid profile.
However, to get more precise results for the spherical bubble,  the local spherical EoM across the bubble wall at steady-state stage needs to be solved, and we leave that to our future study.

\section{Conclusion}
\label{sec:con}
In this work, we have proposed a model-dependent method to calculate the energy budget of the cosmological FOPT.
Taking a representative model as an example, we have illustrated the calculation of the kinetic energy fraction with our method.
By solving the local EoM across the bubble wall, we can simultaneously derive the bubble wall velocity and the boundary conditions of the fluid equations.
Comparing our results with the conventional model-independent method, we have found there are significant differences in the detonation mode.
For the deflagration mode, the differences in the results derived by both methods are less significant.
The origins of these discrepancies are attributed to different EoS and the contribution of the scalar field to the energy-momentum tensor across the bubble wall.
Our model-dependent method gives a more realistic EoS, which is directly derived from the effective potential, and it takes into account the contribution of the scalar field to the energy-momentum tensor.
Hence, our method in principle gives more precise results for the energy budget and the prediction of phase transition GWs, and it could be directly used in other well-motivated particle physics models.

\acknowledgments
X. W. is supported by the China Postdoctoral Science Foundation under Grant No.2022M713642. 
F. P. H. is supported by the National Natural Science Foundation of China (NNSFC) under Grant No. 12205387.

%\newpage
\bibliographystyle{JHEP}
\bibliography{ref.bib}

\end{document}